\documentclass[11pt]{JHEP3}

\usepackage{amsmath}
\usepackage{amsfonts}
\usepackage{mathrsfs}
\usepackage{array}
\usepackage{epsfig}
\usepackage{cite}

\newcommand{\dub}{\,\,\,\,}

\newcommand{\hnab}{\hat{\nabla}}
\newcommand{\hgam}{\hat{\gamma}}

\newcommand\HH{\mathcal{H}}
\newcommand\VV{\mathcal{V}}

\newcommand\MM{\mathcal{M}}

\newcommand\EE{\mathcal{E}}

\newcommand\dM{\partial \MM}

\newcommand\LL{\mathcal{L}}

\newcommand\ZZ{\mathcal{Z}}

\newcommand\eps{\epsilon}

\newcommand\nts{\negthickspace}
\newcommand\bns{\nts \nts \nts}

\newcommand\hg{\hat{g}}
\newcommand\hQ{\hat{Q}}
\newcommand\hU{\hat{U}}
\newcommand\hV{\hat{V}}
\newcommand\hR{\hat{R}}
\newcommand\hK{\hat{K}}
\newcommand\hw{\hat{w}}
\newcommand\hxi{\hat{\xi}}

\newcommand\al{\alpha}
\newcommand\be{\beta}
\newcommand\ga{\gamma}
\newcommand\de{\delta}

\newcommand\defeq{\mathrel{\mathop:}=}

\DeclareMathAlphabet{\mathpzc}{OT1}{pzc}{m}{it}

\newcommand\CurlD{\mathscr{D}}

\newcommand{\field}[1]{\mathbb{#1}}

\newcommand{\BR}{\field{R}}

\newcommand{\arcsinh}{{\rm arcsinh}\,}

\newcommand{\la}{\lambda}

\newcommand{\eq}[2]{\begin{equation} #1 \label{#2} \end{equation}}

\newcommand{\blist}{\begin{itemize}}
\newcommand{\elist}{\end{itemize}}

\newcommand{\killing}{\xi}
\newcommand{\extd}{d}
\newcommand{\mass}{M}

\newcommand{\iscoa}{f}
\newcommand{\iscob}{g}
\newcommand{\iscoc}{h}

\newcommand{\enthalpy}{H}
\newcommand{\isoth}{\kappa_T}

\def\ul{\underline}

\newcommand{\couplingESBH}{\sigma}

\preprint{Brown-HET-1478\\MIT-CTP-3825\\ {\tt hep-th/0703230}}

\title{Thermodynamics of Black Holes in Two (and Higher) Dimensions}

 \author{Daniel Grumiller\\
           Center for Theoretical Physics,
           Massachusetts Institute of Technology,\\
           77 Massachusetts Ave.,
           Cambridge, MA  02139\\
           Email: \email{grumil@lns.mit.edu}}

 \author{Robert McNees\\
          High Energy Theory Group,
          Department of Physics,
          Brown University,\\
          Providence, RI 02912\\
          Email: \email{mcnees@het.brown.edu}}


\abstract{
A comprehensive treatment of black hole thermodynamics in two-dimensional dilaton gravity is presented. We derive an improved action for these theories and construct the Euclidean path integral. An essentially unique boundary counterterm renders the improved action finite on-shell, and its variational properties guarantee that the path integral has a well-defined semi-classical limit.
We give a detailed discussion of the canonical ensemble described by the Euclidean partition function,  and examine various issues related to stability.
Numerous examples are provided, including black hole backgrounds that appear in two dimensional solutions of string theory. 
We show that the Exact String Black Hole is one of the rare cases that admits a consistent thermodynamics without the need for an external thermal reservoir. Our approach can also be applied to certain higher-dimensional black holes, such as Schwarzschild-AdS, Reissner-Nordstr\"om, and BTZ.
}

\begin{document}

\section{Introduction}
\label{sec:Intro}

Two-dimensional gravity provides a useful expedient for testing ideas about quantum gravity in higher dimensions. Technical simplifications in two dimensions often lead to exact results, and it is hoped that this might help to address some of the conceptual problems posed by quantum gravity. At the same time, these simplifications remove many of the features that make gravity interesting. Striking a balance between models that are tractable and models that seem relevant is an art in its own right.


In this paper we study black hole (BH) thermodynamics in two-dimensional models of dilaton gravity. In the absence of matter these models can be solved exactly, but they contain no dynamical degrees of freedom. This seems too restrictive, so we sacrifice integrability and consider models that are coupled to matter. A detailed description of the matter is not necessary for our purposes. We only require the existence of some propagating degrees of freedom, analogous to the radiation emitted by a BH in higher dimensions. We then perform a semi-classical analysis and obtain a comprehensive description of BH thermodynamics for arbitrary models. Despite the fact that we work in two dimensions, these systems exhibit a wide range of interesting thermodynamic phenomena. To some extent this is due to our consideration of `quasilocal' thermodynamic potentials describing BHs in a finite cavity. Such a treatment reveals aspects of BH thermodynamics that are not visible in an asymptotic analysis. Several universal properties of dilaton gravity BHs are identified, as well as structures that are common to families of models. Our approach allows us to study the thermodynamics of backgrounds, both approximate \cite{Witten:1991yr} and exact \cite{Dijkgraaf:1992ba}, associated with two-dimensional solutions of string theory. We are also able to extend our analysis to spherically symmetric BHs in higher dimensions. In particular, we obtain an expression that describes the free energy of both asymptotically flat and asymptotically Anti-de Sitter (AdS) BHs.

Dilaton gravity in two dimensions is conventionally described by the (Euclidean) action
\begin{equation}\label{Action}
  I = - \frac{1}{16\pi G_2}\,\int_{\MM} \nts \nts d^{\,2}x \,\sqrt{g}\, \left[ X\,R - U(X)\,
        \left(\nabla X\right)^2 - 2 \, V(X) \raisebox{12pt}{~}\right]
        -  \frac{1}{8\pi G_2}\, \int_{\dM} \bns dx \, \sqrt{\gamma}\,X\,K ~.
\end{equation}
We define the dilaton $X$ in terms of its coupling to the Ricci scalar by the expression $X\,R$. Different models are distinguished by the kinetic and potential functions $U(X)$ and $V(X)$, cf.~e.g.~\cite{Grumiller:2002nm} for a review. The remaining part of the action is the analog of the Gibbons-Hawking-York boundary term \cite{York:1972sj, Gibbons:1976ue}, where $\gamma_{ab}$ is the induced metric on the boundary and $K$ is the trace of the extrinsic curvature.
Our conventions for curvature tensors and other geometric quantities can be found in appendix \ref{sec:Conventions}. To study the thermodynamics of these theories we employ the Euclidean path integral,
\begin{equation}\label{PartitionFunction}
  \ZZ = \int \CurlD g \CurlD X \, \exp\left(-\frac{1}{\hbar}\,I[g,X] \right) ~.
\end{equation}
The path integral is evaluated by imposing boundary conditions on the fields and then performing the weighted sum over all relevant space-times $(\MM,g)$ and dilaton configurations $X$. In the semi-classical limit it is dominated by contributions from stationary points of the action. This can be verified by expanding it around a classical solution
\begin{equation}\label{Saddle}
  I[g_{cl} + \delta g, X_{cl} + \delta X] = I[g_{cl},X_{cl}] + \de I[g_{cl}, X_{cl}; \delta g, \delta X] + \frac12 \de^2I[g_{cl}, X_{cl}; \delta g, \delta X] + \ldots
\end{equation}
where $\de I$ and $\de^2 I$ are the linear and quadratic terms in the Taylor expansion. The linear term is assumed to vanish on-shell. Then, if the leading term is finite and the quadratic term is positive definite, the saddle point approximation of the path integral is given by
\begin{equation}\label{ApproxPF}
 \ZZ \sim \exp\left(-\frac{1}{\hbar}\,I[g_{cl},X_{cl}]\right) \, \int \CurlD \delta g \CurlD \delta X \, \exp\left(-\frac{1}{2\hbar}\,\de^2I[g_{cl}, X_{cl}; \delta g,\delta X] \right) ~.
\end{equation}
This is interpreted as the partition function in the canonical ensemble, with the temperature set by the path integral boundary conditions.

The transition from \eqref{PartitionFunction} to \eqref{ApproxPF} is complicated by the fact that the action \eqref{Action} does not possess any of the properties listed above. First, the on-shell action diverges. This is familiar from calculations in higher dimensions, and is commonly addressed using a technique known as `background subtraction' \cite{Gibbons:1976ue,Liebl:1997ti}. Second, the linear term in \eqref{Saddle} does not vanish for all field configurations that contribute to the path integral \cite{Regge:1974zd, Mann:2005yr}. This is due to boundary terms whose contribution to $\delta I$ are usually not considered in detail. Schematically, these boundary terms take the form
\begin{equation}\label{ActionVariationIntro}
   \left.\delta I \raisebox{12pt}{\,}\right|_{\rm on-shell} \sim \int_{\dM} \bns dx \, \sqrt{\gamma}\,
    \left[ \pi^{ab} \, \delta \gamma_{ab} + \pi_{X} \, \delta X \, \right] \neq 0 ~.
\end{equation}
Even when we consider models where the boundary conditions imply $\delta \gamma_{ab} \to 0$ and $\delta X \to 0$ at $\dM$, the coefficients of these terms tend to diverge so rapidly that $\delta I$ does not vanish. Finally, the Gaussian integral in \eqref{ApproxPF} is often divergent. In that case the canonical partition function is not well-defined. Instead of describing the thermodynamics of a stable system, \eqref{ApproxPF} yields information about decay rates between various field configurations with the specified boundary conditions \cite{Gross:1982cv}.

We solve the first two problems by replacing \eqref{Action} with an `improved' action $\Gamma$. The improved action is related to \eqref{Action} by a boundary counterterm. The counterterm, which is essentially unique, is obtained by extending the results of \cite{Davis:2004xi} to arbitrary models of dilaton gravity. The improved action is finite on-shell, and $\delta \Gamma$ vanishes for any variation of the fields that preserve the path integral boundary conditions. However, it does not address the remaining problem. The quadratic term $\delta^2 \Gamma$ may result in a Gaussian integral, as in \eqref{ApproxPF}, that is divergent.
This is directly related to the thermodynamic stability of the system. Essentially, the density of states grows so rapidly that the canonical ensemble is not defined. We deal with this in the same manner as York \cite{York:1986it}. The BH is placed inside a cavity that couples the system to a thermal reservoir, with boundary conditions fixed at the wall of the cavity. Following Gibbons and Perry \cite{Gibbons:1992rh}, we identify the location of the wall with the value of a conserved dilaton charge. The canonical ensemble that emerges from this procedure is well defined if and only if the specific heat of the system is positive. Typically this requires a finite cavity, but there are some notable exceptions. Having resolved the three problems described above, the approximation \eqref{ApproxPF}, with $\Gamma$ replacing $I$, gives the canonical partition function for the system.

We proceed as follows: in section \ref{sec:SPI} we review BH solutions of two-dimensional dilaton gravity \cite{Frolov:1992xx, Klosch:1996fi}, and examine the action \eqref{Action} and its variation evaluated on these backgrounds. We then derive the improved action $\Gamma$ and verify its properties. The Euclidean path integral is used to define the canonical partition function in section \ref{sec:Thermo}, from which all of our results on thermodynamics follow. We obtain expressions for thermodynamic quantities in an arbitrary model of two-dimensional dilaton gravity, establish universal results like the first law of BH thermodynamics, and examine several issues related to stability. In section \ref{sec:2DExamples} these results are applied to two significant families of models. Section \ref{sec:StringExamples} addresses BHs that appear in models obtained from string theory. This includes a treatment of the Exact String BH \cite{Dijkgraaf:1992ba}. The results obtained in section \ref{sec:Thermo} can also be applied, in certain cases, to BH solutions of higher dimensional theories. This is discussed in section \ref{sec:HigherDimensionalExamples}, where we study spherically symmetric BHs with or without a cosmological constant, and the toroidal Kaluza-Klein reduction of the BTZ BH \cite{Banados:1992wn}. 
Finally, in section \ref{sec:Discussion} we summarize our results, discuss further applications, and suggest possible generalizations. A collection of appendices contain our conventions, a review of various models, and additional technical results.

Before proceeding to section \ref{sec:SPI}, we point out a few important conventions. Euclidean signature is used throughout this paper. Nevertheless, we frequently use terms such as `Minkowski' and `horizon' as if we were working in Lorentzian signature. The dimensionless Newton's constant that appears in the action is set to $8\pi G_2 = 1$. It is restored in certain expressions, when necessary.

\section{Semi-Classical Approximation of the Dilaton Gravity Path Integral}
\label{sec:SPI}

In this section we consider the Euclidean path integral formulation of dilaton gravity in two dimensions. Our analysis shows that the standard form of the action is not stationary under the full class of field variations that preserve the boundary conditions. We construct an improved action that solves this problem, and obtain the semi-classical approximation of the corresponding path integral.

\subsection{Equations of Motion}
\label{sec:Action}

The equations of motion are obtained by extremizing the action \eqref{Action} with respect to small variations of the fields that vanish at the boundary. This yields
\begin{gather} \label{MetricEOM}
     U(X)\,\nabla_{\mu}X \nabla_{\nu}X - \frac{1}{2}\,g_{\mu\nu} U(X) (\nabla X)^2
        - g_{\mu\nu} V(X) + \nabla_{\mu} \nabla_{\nu} X - g_{\mu\nu} \nabla^{2} X  = 0 \\ \label{XEOM}
     R + \partial_X U(X) (\nabla X)^2 + 2 \,U(X) \nabla^{2} X - 2 \,\partial_X V(X)  = 0 ~.
\end{gather}
Solutions of these equations always possess at least one Killing vector whose orbits are curves of constant $X$ \cite{Schmidt:1989ws, Banks:1990mk}. We fix the gauge so that the metric is diagonal. The solutions
\eq{
   X = X(r)\,, \qquad ds^2 = \xi(r) \,d\tau^2 + \frac{1}{\xi(r)}\,dr^2\,,
}{metric}
with
\begin{eqnarray}
    \partial_r X & = & e^{-Q(X)} \label{XPrimeDef} \\ \label{xiDef}
    \xi(X) & = & w(X) \, e^{Q(X)}\,\left( 1 - \frac{2\,M}{w(X)} \right)
\end{eqnarray}
are expressed in terms of two model-dependent functions,
\begin{eqnarray}\label{QDef}
    Q(X) & \defeq & Q_0 \, + \int^{X} \bns d\tilde{X} \, U(\tilde{X})\,, \\ \label{wDef}
        w(X) &  \defeq & w_0 -2 \, \int^{X} \bns d\tilde{X} \, V(\tilde{X}) \, e^{Q(\tilde{X})}\,.
\end{eqnarray}
Here $Q_0$ and $w_0$ are constants, and the integrals are evaluated at $X$. Notice that $w_0$ and the integration constant $M$ contribute to $\xi(X)$ in the same manner. Together, they represent a single parameter that has been partially incorporated into the definition of $w(X)$. By definition they transform as $w_0 \to e^{\Delta Q_0} w_0$ and $M \to e^{\Delta Q_0} M$ under the shift $Q_0 \to Q_0 + \Delta Q_0$. This ensures that the functions \eqref{XPrimeDef} and \eqref{xiDef} transform homogeneously, allowing $Q_0$ to be absorbed into a rescaling of the coordinates. Therefore, the solution is parameterized by a single constant of integration. With an appropriate choice of $w_0$ we can restrict $M$ to take values in the range $M \geq 0$ for physical solutions. As evident from \eqref{metric} the Killing vector described above is $\partial_{\tau}$. For brevity we refer to $\sqrt{\xi(X)}$ from \eqref{xiDef} as the ``Killing norm''. If it vanishes we encounter a Killing horizon. The square of the Killing norm for the solution with $M=0$ is
\begin{equation}\label{GroundState}
  \xi_{0}(X) \defeq w(X) \, e^{Q(X)} ~.
\end{equation}
This solution, which we often refer to as the `ground state', plays an important role throughout this paper~\footnote{If the function $V(X)$ has one or more zeroes at finite values of $X$ there is a second, inequivalent family of solutions with constant dilaton. In that case, the identification of the actual ground state of the model becomes more involved. This issue will be addressed in subsection \ref{sec:CDV}. }.

\subsection{Black Holes}
\label{sec:BlackHoles}

The constant of integration that labels solutions of the equations of motion is analogous to the mass parameter that appears in the Schwarzschild metric. This constant has been partially absorbed into the definition of $w(X)$ in such a way that $M=0$ represents the ground state of a particular model. We refer to solutions with $M > 0$, which may exhibit horizons, as BHs.

In the models we consider the metric function $\xi(X)$ is non-negative over a semi-infinite interval
\begin{equation}\label{Interval}
X_h \leq X < \infty ~.
\end{equation}
The upper end of this interval corresponds to the asymptotic region of the space-time. The function $w(X)$ generally diverges in this limit
\begin{equation}\label{wAsymptotic}
   \lim_{X \to \infty} w(X) \to  \infty
\end{equation}
so that the asymptotic behavior of the metric is typified by the ground state solution \eqref{GroundState}. The lower end of the interval \eqref{Interval} is either the origin, or a Killing horizon. Assuming that the function $e^{Q(X)}$ is non-zero for finite values of $X$, the location of the horizon is related to the parameter $M$ by
\begin{equation}\label{Horizon}
    w(X_h) = 2\,M  ~.
\end{equation}
If this condition admits multiple solutions then $X_h$ is taken to be the outermost horizon, so that $w(X) > 2M$ for $X > X_h$.

Regularity of the metric at the horizon fixes the periodicity $\tau \sim \tau + \beta$ of the Euclidean time, which is given by
\begin{equation}\label{beta}
   \beta = \left. \frac{4\pi}{\partial_r \xi}\, \right|_{r_h}
    = \left. \frac{4\pi}{w'(X)} \,\right|_{X_h} ~.
\end{equation}
The inverse periodicity is related to the surface gravity of the BH by $2\pi \beta^{-1} = \kappa$.
In an asymptotically flat space-time, $\beta^{-1}$ is also the temperature measured by an observer at infinity. In a slight abuse of notation, we denote this quantity by $T$
\begin{equation}\label{T}
  T = \beta^{-1} =  \left. \frac{w'(X)}{4\,\pi}  \right|_{X_h}~.
\end{equation}
This should not be confused with the proper local temperature
, which is related to $\beta^{-1}$ by a position dependent red-shift (the `Tolman factor'\cite{Tolman:1934})
\begin{equation}\label{Tc}
  T(X) =  \frac{1}{\sqrt{\xi(X)}} \,\beta^{-1} ~.
\end{equation}
The proper temperature at infinity coincides with \eqref{T} only if $\xi(X) \to 1$ as $X \to \infty$.

At this point it is instructive to evaluate the action for the BH solution, which we expect is related to the semi-classical approximation of the BH partition function. The bulk terms in \eqref{Action} are straightforward, but we need to specify how the boundary is treated in order to calculate the Gibbons-Hawking-York term. It is apparent from \eqref{metric} that the only component of $\dM$ that contributes to this calculation is a surface orthogonal to the coordinate $r$. Depending on the choice of coordinate system this may occur either at some finite value of $r$, or in a limit such as $r \to \infty$. In either case, the boundary is assumed to correspond to $X \to \infty$, which means that the boundary terms in the action should be calculated via a suitable limiting procedure. We implement this by placing a regulator $X \leq X_{\rm reg}$ on the dilaton, and treating the isosurface $X=X_{\rm reg}$ as the boundary. Then the on-shell action is
\begin{equation}\label{DivergentAction}
  I_{\rm reg} = \beta \, \left(\raisebox{12pt}{} 2\,M - w(X_{\rm reg})  - 2\,\pi X_{h}\,T \right) ~.
\end{equation}
The regulator should be removed by taking the limit $X_{\rm reg} \to \infty$, which recovers the full BH space-time. At this point we encounter a problem: the asymptotic behavior \eqref{wAsymptotic} of $w(X_{\rm reg})$ causes the action to diverge. How should this term be addressed? One possibility is that it should be discarded, and the finite remainder then interpreted as the action for the space-time. This strategy is unsatisfactory for two reasons. First, it fails to reproduce the correct thermodynamics of the BH. Second, as we shall demonstrate in the next subsection, the variational properties of the action \eqref{Action} are not consistent with our assumptions about the path integral. Neither issue is resolved by the sort of ad-hoc subtraction described above.

\subsection{Variational Properties of the Action}
\label{sec:Variation}

The equations of motion were described in section \ref{sec:Action} as the conditions for extremizing the action. In this section we show that this is true for `localized' variations of the fields, but not for generic variations that preserve the boundary conditions. This has important consequences for the path integral formulation of the theory. We begin by considering small, independent variations of the metric and dilaton. The corresponding change in the action is
\begin{equation}\label{ActionVariation}
   \delta I = - \frac{1}{2} \, \int_{\MM}\nts d^{\,2}x \, \sqrt{g} \, \left[ \raisebox{12pt}{~}
    \EE^{\mu\nu} \, \delta g_{\mu\nu} + \EE_X \, \delta X \,\right]
    + \int_{\dM} \bns dx \, \sqrt{\gamma}\,\left[ \raisebox{12pt}{~} \pi^{ab} \, \delta \gamma_{ab}
    + \pi_{X} \, \delta X \, \right] ~.
\end{equation}
The terms $\EE_{\mu\nu}$ and $\EE_X$ in the bulk integral are the left-hand-sides of \eqref{MetricEOM} and \eqref{XEOM}, respectively. In addition to the bulk term, there is also a boundary term that depends on  momenta conjugate to $\gamma_{ab}$ and $X$
\begin{equation}\label{Momenta}
         \pi^{ab} = - \frac{1}{2} \, \gamma^{ab} \, n^{\mu} \nabla_{\mu} X\,,\qquad
    \pi_{X}   =  \, U(X) \, n^{\mu} \nabla_{\mu} X - \, K ~.
\end{equation}
These momenta are defined with respect to evolution of the fields along the direction normal to the boundary, indicated here by the vector $n^{\mu}$. When evaluated on a solution of the classical equations of motion, the change in the action \eqref{ActionVariation} should vanish for any variation of the fields that preserves the boundary conditions. A direct calculation, which we shall carry out below, shows that this is not the case. Instead, we find that $\delta I \neq 0$ unless we restrict our attention to localized variations of the fields. In this context a field variation is `localized' if its asymptotic behavior satisfies some condition that is more restrictive than simply preserving the boundary conditions. Essentially, the change in the action only vanishes for variations that fall off more rapidly at infinity than required by the boundary conditions.

The claim that $\delta I$ does not vanish for physically reasonable variations of the fields may seem odd. The Gibbons-Hawking-York term in \eqref{Action} is usually motivated by stating that it leads to an action with a ``well defined" variational principle. But the Gibbons-Hawking-York term only ensures that fields obey Dirichlet conditions at $\dM$. It does not guarantee that the boundary term in $\delta I$ vanishes for arbitrary $\delta \gamma_{ab}$ and $\delta X$ that preserve these boundary conditions. This issue has been studied in higher dimensions for gravitational theories with both asymptotically flat \cite{Regge:1974zd, Mann:2005yr} and asymptotically AdS \cite{Papadimitriou:2005ii, McNees:2005my} boundary conditions. It is especially important for the path integral formulation of the theory. Given a solution $(\xi_{cl},X_{cl})$ of the equations of motion, the path integral also receives contributions from field configurations whose asymptotic behavior is $(\xi_{cl}+\delta\xi,X_{cl}+\delta X)$, where $\delta \xi$ and $\delta X$ are generic variations consistent with the boundary conditions. The $\hbar \to 0$ limit of the path integral is dominated by solutions of the classical equations of motion if, and only if, $\delta I$ vanishes for all such variations.

We show now that the action \eqref{Action} does not satisfy this condition. Evaluating \eqref{ActionVariation} for a solution of the form \eqref{metric} gives
\begin{equation}\label{dAction0}
  \delta I_{\rm reg} =  \int d\tau \, \left[ -\frac{1}{2} \, \partial_r X
    \, \delta \xi  +  \, \left(U(X) \, \xi(X) \, \partial_r X
    - \frac{1}{2}\,\partial_r \xi\right) \, \delta X \right] ~.
\end{equation}
The subscript on $\delta I$ indicates that the boundary term must be evaluated using a regulator, as in the previous subsection. The boundary term should vanish as the regulator is removed to infinity. To show that this is not the case, it is enough to set $\delta X = 0$ and consider the variation $\delta \xi$. Using \eqref{XPrimeDef}, the coefficient of $\delta \xi$ is proportional to $e^{-Q(X)}$. Depending on the model, this term may either diverge, approach a constant, or vanish at $\dM$. Now we need to specify how a generic variation $\delta \xi$ is allowed to behave at $\dM$. The models we consider allow for a wide range of asymptotic behavior. In particular, $\xi$ might diverge in the $X \to \infty$ limit. Thus, we cannot assume that $\delta \xi$ vanishes at $\dM$, as we would if $\xi$ approached a constant at $\dM$. Instead we must appeal to the general solution \eqref{xiDef}, from which we infer the asymptotic behavior of a $\delta \xi$ that preserves the boundary conditions.  Recall that the leading asymptotic behavior of $\xi$ is determined by $\xi_0 = w e^{Q}$, since $w \to \infty$ at $X \to \infty$. We assume that the boundary conditions are preserved by variations whose asymptotic behavior is similar to the sub-leading term in \eqref{xiDef}. This means that we must consider the change in the action due to
\begin{equation}\label{xiVariation}
    \delta \xi = \delta M \, e^{Q(X)}
\end{equation}
where $\delta M$ is an arbitrary infinitesimal constant. The corresponding change in the action is non-zero and independent of the regulator
\begin{equation}\label{deltaI}
  \delta I =   \,  \int d\tau \, \delta M  \, \neq \, 0 ~.
\end{equation}
This shows that solutions of \eqref{MetricEOM} and \eqref{XEOM} do not extremize the action \eqref{Action} against generic variations $\delta \xi$ that preserve the boundary conditions on $\xi$. In appendix \ref{sec:NewVariables} we repeat this analysis in terms of a new metric variable $\hxi$ that obeys the standard Dirichlet condition $\hxi = 1$ at $\dM$. Then, in appendix \ref{sec:GaugeIndependence}, we show that the result \eqref{deltaI} does not depend on the diagonal gauge used in \eqref{metric}.

This explains why one cannot drop the regulator-dependent term in \eqref{DivergentAction} and assign a meaning to the part that remains. The semi-classical approximation of the path integral is only well-defined if $\delta I = 0$ for the full class of field variations encountered in the path integral.
With this in mind, we look for an improved action that is both finite and stationary for solutions of the classical equations of motion.

\subsection{The Hamilton-Jacobi Boundary Term}
\label{sec:BoundaryTerm}

In the previous section we found that the variational properties of the action are not compatible with the semi-classical approximation of the path integral. The same problem was encountered in \cite{Davis:2004xi}, where the authors studied BH solutions of the tree-level equations for type 0A string theory in two dimensions. In this section we generalize the techniques used in \cite{Davis:2004xi} to arbitrary theories of dilaton gravity. The result is a new action $\Gamma$ that is related to $I$ by
\begin{equation}\label{GammxiDef}
   I = \Gamma + I_{CT} ~.
\end{equation}
The term $I_{CT}$ is a boundary integral of the form
\begin{equation}\label{CT1}
  I_{CT} = \int_{\dM} \bns dx \, L_{CT}(\gamma,X) ~.
\end{equation}
The integrand may depend on the fields at the boundary, and their derivatives along the
boundary, but not their normal derivatives. This ensures that $\Gamma$ and $I$ yield the same
equations of motion. The term $I_{CT}$ is then defined so that solutions of these equations are
stationary points of $\Gamma$, with respect to any variation of the fields that preserves the
boundary conditions. This construction is similar to the boundary counterterm technique used in
the AdS/CFT correspondence.  Although this approach was originally devised to cancel
divergences in the action \cite{Henningson:1998gx, Balasubramanian:1999re, Emparan:1999pm,
deHaro:2000xn}, it was later shown that the boundary counterterms follow from the requirement
that the action admits a well-defined variational problem with asymptotically AdS boundary
conditions \cite{Papadimitriou:2005ii}. A similar modification of the Einstein-Hilbert action
was proposed in \cite{Mann:2005yr, Mann:2006bd, Astefanesei:2006zd}, where the authors
considered space-times with asymptotically flat boundary conditions.

Following the derivation in \cite{Davis:2004xi}, the boundary counterterm is identified with a solution of the Hamilton-Jacobi equation for the on-shell action. This equation can be derived by starting with the Hamiltonian constraint that follows from the action \eqref{Action}
\begin{equation}\label{Hamiltonian}
   \HH = 2 \, \pi_{X} \, \gamma_{ab} \, \pi^{ab} + 2 \, U(X) \, \left( \gamma_{ab} \,
    \pi^{ab} \right)^2 + \,V(X)  = 0 ~.
\end{equation}
Recall that the momenta, defined in \eqref{Momenta}, appear as boundary terms in the variation of the action. When $\delta I$ is evaluated for a solution of the equations of motion the bulk term vanishes, leaving only the boundary term. Thus, the momenta can be expressed as functional derivatives of the \emph{on-shell} action
\begin{equation} \label{FDMomenta}
   \pi^{ab} = \frac{1}{\sqrt{\gamma}}\,\frac{\delta}{\delta \gamma_{ab}}\left(
    \left.  I \raisebox{12pt}{~}\right|_{\EE=0}\right) \quad \quad \quad
    \pi_{X} = \frac{1}{\sqrt{\gamma}}\,\frac{\delta}{\delta X}\left(
    \left.  I \raisebox{12pt}{~}\right|_{\EE=0}\right) ~.
\end{equation}
Rewriting the Hamiltonian constraint \eqref{Hamiltonian} in terms of these expressions gives a non-linear functional differential equation for the on-shell action which is difficult to solve.
But in our case symmetry arguments reduce the problem to a linear differential equation for a function of $X$, as we now demonstrate. The solution $I_{CT}$ should be invariant under diffeomorphisms of $\dM$, so we express its integrand as $L_{CT}(\gamma,X) = \sqrt{\gamma}\,\LL_{CT}(\gamma,X)$. Because there are no other intrinsic, diffeomorphism covariant scalars constructed from $\gamma$ in one dimension, it follows that $\LL_{CT}$ can depend only on $X$ and its derivatives along the boundary. But terms involving derivatives can be ignored, because the boundary is an isosurface of $X$. Thus, the scalar density $\LL_{CT}$ depends only on $X$ and the boundary counterterm takes the form
\begin{equation}
   I_{CT} =  \, \int_{\dM} \bns d\tau \, \sqrt{\gamma} \LL_{CT}(X) ~.
\end{equation}
The functional derivatives \eqref{FDMomenta} applied to this integral give `momenta' that we denote
\begin{equation}\label{GammaMomenta}
  p^{ab} = \frac{1}{2} \, \gamma^{ab} \, \LL_{CT}(X) \quad \quad \quad
  p_{X} =  \, \partial_{X} \LL_{CT}(X)
\end{equation}
to distinguish them from $\pi^{ab}$ and $\pi_{X}$. Applying these expressions in \eqref{Hamiltonian}, the Hamiltonian constraint becomes a linear differential equation for $\LL_{CT}^2$,
\begin{equation}\label{HJEqn2}
  \frac{1}{2} \, \partial_{X} \left(\LL_{CT}(X)^2 \right) + \frac{1}{2} \, U(X) \, \LL_{CT}(X)^2 + V(X) = 0 ~.
\end{equation}
The general solution of this equation is
\begin{equation}\label{LCT}
  \LL_{CT}(X) =  - \sqrt{e^{-Q(X)}\,\left(w(X) + c\right)}
\end{equation}
where $Q(X)$ and $w(X)$ were defined in \eqref{QDef} and \eqref{wDef}, and $c$ is an arbitrary constant. Note that the Hamilton-Jacobi equation determines $(\LL_{CT})^{\,2}$, and the negative root has been chosen in \eqref{LCT}. With the definition \eqref{GammxiDef}, this root leads to an action $\Gamma$ with the correct properties.

In many models there are natural arguments for setting the constant $c$ to zero. For instance, in models associated with two-dimensional string theories the action $I$ is invariant under Buscher duality \cite{Buscher:1987sk}. 
The action $\Gamma$ is expected to possess the same invariances as $I$, and this is only the case if $c = 0$, as pointed out in \cite{Davis:2004xi}. More generally, setting $c=0$ guarantees that $\LL_{CT}$ is independent of the scaling ambiguity associated with the constant $Q_0$ that appears in \eqref{QDef}. Therefore, we set $c = 0$ and obtain the boundary counterterm
\begin{equation}\label{ICT}
  I_{CT} = -  \int_{\dM} \bns dx \,\sqrt{\gamma} \, \sqrt{w(X) e^{-Q(X)}} ~.
\end{equation}
This leads to the `improved' action
\begin{equation}\label{GammxiDef2}
  \Gamma = I +  \int_{\dM} \bns dx \,\sqrt{\gamma} \, \sqrt{w(X) e^{-Q(X)}} ~.
\end{equation}

\subsection{Path Integral Based on the Improved Action}
\label{sec:FinalAction}

Based on the analysis of the previous subsections, we conclude that the Euclidean path integral for two-dimensional dilaton gravity should weight field configurations using the exponential of the improved action,
\begin{equation}\label{PartitionFunction2}
  \ZZ  = \int \CurlD g \CurlD X \, \exp\left(- \frac{1}{\hbar}\,\Gamma[\,g,X]\raisebox{10pt}{~} \right) \,,
\end{equation}
rather than \eqref{PartitionFunction}.
In this section we justify this claim by establishing two important properties of $\Gamma$. First we show that solutions of \eqref{MetricEOM} and \eqref{XEOM} are stationary points of \eqref{GammxiDef2} with respect to variations of the fields that preserve the boundary conditions. This means that the semi-classical approximation of \eqref{PartitionFunction2} is dominated by solutions of the classical equations of motion \eqref{MetricEOM} and \eqref{XEOM}. Second, we evaluate the action for the BH space-times discussed in section \ref{sec:BlackHoles} and show that the result is finite.

The variation of the action $\Gamma$ is
\begin{equation}\begin{split}
\label{dGamma}
  \delta \, \Gamma = &{} - \frac{1}{2} \, \int_{\MM}\nts d^{\,2}x \, \sqrt{g} \, \left[
    \raisebox{12pt}{~} \EE^{\mu\nu} \, \delta g_{\mu\nu} + \EE_X \, \delta X \,\right] \\
  &{} + \int_{\dM} \bns dx \, \sqrt{\gamma}\,\left[ \raisebox{12pt}{~} (\pi^{ab} - p^{ab})
    \, \delta \gamma_{ab} + (\pi_{X}-p_X) \, \delta X \, \right] ~.
\end{split}
\end{equation}
This differs from \eqref{ActionVariation} by boundary terms coming from the variation of \eqref{ICT}. We can repeat the analysis of section \ref{sec:Variation}, evaluating $\delta \Gamma$ on-shell and checking that the boundary term vanishes for appropriate variations of the metric and dilaton. The terms involving $\pi^{ab}$ and $\pi_X$ were evaluated in \eqref{dAction0}. Using \eqref{GammaMomenta}, the contributions due to $I_{CT}$ are
\eq{
  p^{ab} = - \frac{1}{2} \, \gamma^{ab} \, e^{-Q(X)}\,\sqrt{\xi_{0}(X)}\,,\qquad 
  p_{X} =  \frac{1}{2} \, e^{-Q(X)}\,U(X)\,\sqrt{\xi_{0}(X)}
     - \frac{1}{2} \, \frac{w'(X)}{\sqrt{\xi_{0}(X)}}~.
}{eq:test}
As before, the boundary term is evaluated at the cut-off $X=X_{\rm reg}$
\begin{multline}
   \delta \,\Gamma_{\rm reg} = \frac{1}{2} \, \int d\tau  \, \Bigg[ e^{-Q(X)} \, \Bigg(\sqrt{\frac{\xi_{0}(X)}{\xi(X)}}-1\Bigg) \, \delta \xi \\
 + \Bigg( \xi(X) \,U(X)\,e^{-Q(X)}\,\Big(1-\sqrt{\frac{\xi_{0}(X)}{\xi(X)}}\Big) - w'(X) \, \Big(1-\sqrt{\frac{\xi(X)}{\xi_{0}(X)}}\Big)\Bigg) \,\delta X \Bigg]~. \label{dGammaAlso}
 \end{multline}
This result simplifies a great deal when the regulator is moved far into the asymptotic region of the space-time. The function $w(X)$ becomes very large in this limit, and the coefficients of $\delta \xi$ and $\delta X$ can be expanded in powers of $1/w$. The leading order terms in this expansion are
\begin{equation}\label{dGamma2}
   \delta \, \Gamma_{\rm reg} = \frac{M}{2} \, \int d\tau \, \left[ \frac{1}{\xi_{0}(X)}\,
      \delta \xi - \left(U(X) + \frac{w'(X)}{w(X)}\right)\,\delta X\right] + \dots
\end{equation}
The variation of the metric $\delta \xi$ behaves as \eqref{xiVariation}, so the first term clearly vanishes in the $X_{\rm reg} \to \infty$ limit. However, the analysis of \eqref{dGamma2} is much simpler if we use the change of variables $g_{\mu\nu} = e^{2\sigma(X)}\,\hg_{\mu\nu}$ defined in \eqref{newmetric}, with $\sigma = w(X)e^{Q(X)}$ as in \eqref{eq:apB3}. Then the variation of $\Gamma$ is given by \eqref{dGammaAlso} with all un-hatted quantities replaced by their hatted equivalents. Using \eqref{NewFunctions}, the leading terms in the large-$w$ expansion of $\delta\Gamma$ are
\begin{equation}
 \lim_{X_{\rm reg} \to \infty} \delta \Gamma_{\rm reg} = \frac{M}{2} \lim_{X_{\rm reg} \to \infty}  \int d\tau \, \left[ \delta \hxi
    + M \, \delta \Big(\frac{1}{w}\Big) \right] ~.
\end{equation}
The new variable $\hxi$ obeys the standard Dirichlet condition $\hxi = 1$ at the boundary, which means that $\delta \hxi$ vanishes as $X_{\rm reg} \to \infty$. The second term must also vanish in this limit since we assume that $w \to \infty$ asymptotically. Thus we have established
\begin{equation}\label{equationwithoutlabel}
  \lim_{X_{\rm reg} \to \infty} \delta \,\Gamma_{\rm reg} = 0 ~.
\end{equation}
We conclude that solutions of the classical equations of motion are stationary points of $\Gamma$ with respect to all variations of the fields that preserve the boundary conditions.

Now we evaluate the action for the BH solution. In section \ref{sec:BlackHoles} we found that the regulated on-shell action $I_{\rm reg}$ contains a term that diverges in the $X_{\rm reg} \to \infty$ limit. Including the contribution from the boundary counterterm, the improved action is
\begin{equation}\label{CutOffAction}
  \Gamma_{\rm reg}  =  \, \beta \, \left(\raisebox{12pt}{} w(X_{\rm reg}) \, \sqrt{1 - \frac{2 M}{w(X_{\rm reg})}} - w(X_{\rm reg}) + 2\,M - 2\,\pi X_{H}\,T \right) ~.
\end{equation}
The regulator can be safely removed to obtain a finite result
\begin{equation}\label{FinalFiniteAction}
  \lim_{X_{\rm reg} \to \infty} \Gamma_{\rm reg} = \,\beta\, \left( \raisebox{10pt}{}M - 2\pi X_H \, T\right) ~.
\end{equation}
Notably, this differs by an amount $\beta\,M$ from the part of \eqref{DivergentAction} that was independent of the regulator.

\section{Thermodynamics in the Canonical Ensemble}
\label{sec:Thermo}

Motivated by York's path integral formulation of gravitational thermodynamics \cite{York:1986it}, we now introduce an upper bound $X_c$ on the dilaton. This constrains the dilaton to a `cavity' $X \leq X_c$ whose wall is the dilaton isosurface $X=X_c$. The cavity is in contact with a thermal reservoir that maintains boundary conditions at $X_c$~\footnote{Note that $X_c$ has a clear physical interpretation, unlike the regulator $X_{\rm reg}$ introduced in the previous section. The regulator appeared as an ingredient of the limiting procedure used to evaluate boundary quantities. As such, the limit $X_{\rm reg} \to \infty$ was always assumed to be the final step of any calculation. This is not the case with $X_c$, which may remain finite.}. The semi-classical approximation of the corresponding path integral yields the partition function in the canonical ensemble. The consistency of this interpretation rests upon the condition of thermodynamic stability, which we discuss in subsections \ref{sec:specificheat} and \ref{sec:Stability}.

\subsection{Dilaton Charge and Free Energy}

In standard thermodynamics the canonical ensemble fixes the temperature and volume of a system. The local temperature $T_c$ at the cavity wall is fixed by requiring the proper periodicity of the Euclidean time to be $\beta_c$. But the volume of the cavity at constant Euclidean time is not a well-defined quantity, due to the possibility of a horizon $X_h < X_c$. In higher dimensions this issue is addressed by holding the area of the cavity wall fixed, instead \cite{York:1986it}. We essentially do the same, but in an indirect way. In two-dimensional dilaton gravity any sufficiently regular function of the dilaton can be used to construct a conserved current. As a result, there are an infinite number of conserved charges associated with the dilaton. Following Gibbons and Perry \cite{Gibbons:1992rh}, we choose a particular conserved charge that we denote by $D(X)$ and hold this quantity fixed at $X_c$.  One convenient choice is
\begin{equation}\label{DilaCharge}
  D_c = X_c \, ,
\end{equation}
which plays a role similar to the area of the cavity wall in higher dimensions. This choice can be regarded as ``natural" as it is precisely the quantity that couples to the Ricci scalar and the extrinsic curvature in the action. Another ``natural'' possibility for the dilaton charge,
\eq{
\bar{D}_c = e^{-Q_c}\sqrt{\killing_0(X_c)}\,,
}{eq:dc1}
coincides with the solution of the Hamilton-Jacobi equation \eqref{ICT}. We shall discuss in detail how various thermodynamical properties depend on the choice of dilaton charge.

Fixing $D_c$ and $\beta_c$ specifies the boundary conditions for the path integral. The classical solutions consistent with these boundary conditions correspond to real, non-negative values of $\mass$ that satisfy the condition
\begin{equation}\label{ClassicalSolutions}
  \beta_c = \sqrt{\xi(D_c, \mass)} \, \beta(M) ~,
\end{equation}
where $\beta(M)$ is the periodicity of the Euclidean time given in \eqref{beta}.
In general, there may be zero, one, or multiple classical solutions that satisfy \eqref{ClassicalSolutions}. Issues related to the existence of multiple solutions will be discussed in subsection \ref{sec:Stability}. For the next several subsections, we assume that we are working with a single solution. In that case the Helmholtz free energy of the solution is related to the on-shell action $\Gamma_c$ by
\begin{equation}\label{ZFRelation}
  F_c (T_c, D_c) = T_c \, \Gamma_c(T_c,D_c) \,  ~.
\end{equation}
From \eqref{Tc}, \eqref{CutOffAction} and \eqref{ZFRelation} we obtain
\begin{equation}\label{FreeEnergy}
  F_c = - 2\pi\,X_h \, T_c + e^{-Q_c}\,\left(\sqrt{\killing_0}-\sqrt{\killing_c}\right) ~.
\end{equation}
It is worth mentioning that the free energy for the ground state solution vanishes if either $T=0$ or $X_h = 0$. This statement does not depend on the location of the cavity wall.

The interpretation of \eqref{ZFRelation} as the Helmholtz free energy assumes thermodynamic stability. The consistency of this assumption will be addressed in subsections \ref{sec:specificheat} and \ref{sec:Stability}. In most models thermodynamic stability requires a \emph{finite} cavity. However, we are often interested in the properties of an isolated BH without an enclosing cavity. This is accomplished by taking the limit $X_c \to \infty$, while simultaneously changing $\beta_c$ according to \eqref{ClassicalSolutions}. This limit, if it exists, can be thought of as removing the cavity wall to infinity while adiabatically relaxing the temperature to preserve $\mass$. Motivated by curiosity, we shall sometimes apply this limit to the result of a calculation without worrying about whether or not it is consistent with thermodynamic stability (usually it is not). We emphasize that, because the on-shell action is finite in this limit, any singular behavior in $F_c$ as $X_c \to \infty$ is due to the Tolman factor in $T_c$, as evident from \eqref{ZFRelation}.

Another issue related to stability concerns the possibility of decay into a field configuration with a smaller free energy. The BH solution is unstable against decay into any field configuration (that obeys the boundary conditions) with a free energy $F_g$ smaller than $F_c$ from \eqref{FreeEnergy}. We shall discuss this issue in more detail later on. For now, we simply point out that there are usually relevant field configurations with a non-positive free energy, so a minimal condition for stability of the BH solution is $F_c\leq 0$. It implies
\eq{
2M-2\pi X_h T \leq w_c \left(1-\sqrt{1-\frac{2M}{w_c}}\right) ~.
}{eq:HFS}
This condition is always fulfilled for boundary conditions that admit solutions with horizons close to the cavity wall: $w_c = 2M + \eps$. As we shall see in later parts of this section, the inequality \eqref{eq:HFS} is usually not fulfilled for solutions \eqref{ClassicalSolutions} with $2M \ll w_c$.

Keeping in mind the caveats described above, the collection of formulas in appendix \ref{sec:Formulas} can be applied to the Helmholtz free energy \eqref{FreeEnergy} to calculate all thermodynamical quantities. We start with the most prominent one.

\subsection{Entropy and Area}
\label{sec:Entropy}

The entropy is given by
\begin{equation}\label{entropy}
  S_c = - \left. \frac{\partial F_c}{\partial T_c} \right|_{D_c}\,.
\end{equation}
The variation of $F_c$ at fixed $D_c$ can be expressed as
\begin{equation}\label{}
        d F_c \Big|_{D_c}
\, = - 2\pi\,X_h \, dT_c + \,\left( \frac{1}{\sqrt{\killing_c}} - \frac{T_c}{T}
        \right) \, d\mass ~.
\end{equation}
The last term, which was obtained using \eqref{dX_and_dT} and \eqref{dA_D}, vanishes by the defining relation \eqref{Tc}. Thus, we find the well-known result \cite{Gibbons:1992rh, Nappi:1992as, Gegenberg:1994pv, Davis:2004xi, Grumiller:2006rc}
\begin{equation}\label{Entropy2}
  S = 2\pi\,X_h ~.
\end{equation}
The entropy is independent of $X_c$, and depends only on the value of the dilaton at the horizon. As expected entropy is exclusively a property of the horizon.  Furthermore, this result is universal and applies to any model \eqref{Action} regardless of the potentials $U$ and $V$. Restoring the two-dimensional Newton's constant $G_2$ the result \eqref{Entropy2} becomes $S=X_h/(4G_2)$. It is useful to introduce the notion of area at this point in order to compare with the standard Bekenstein-Hawking formula. The area of a sphere of radius $r$ embedded in $d$ spatial dimensions is given by $A_d=2\pi^{d/2}r^{d-1}/\Gamma(d/2)$. In the limit $d\to 1$ the area becomes independent of the radius and the ``sphere" consists of two disjoint points. It is suggestive to restrict to a connected component and therefore to associate only one of these two points with the horizon. Thus we define $A_{h}=A_1/2=1$. The effective Newton coupling in a scalar-tensor theory is given by $G_{\rm eff} = G_2/X$, where $X$ is either approximately constant or evaluated at some characteristic scale. The only scale available here is provided by the horizon, so we define $G_{\rm eff}=G_2/X_h$. Then \eqref{Entropy2} is equivalent to the Bekenstein-Hawking relation
\begin{equation}\label{eq:BH}
  S = \frac{A_{h}}{4\,G_{\rm eff}}\,,
\end{equation}
so that the entropy is one quarter of the horizon area in `effective Planck units'. In section \ref{sec:HigherDimensionalExamples}, when we consider the reduction of a spherically symmetric BH from $d+1>2$ dimensions down to two dimensions, we shall find that the conserved charge $D_c$ always corresponds to the higher-dimensional area, in $d+1$ dimensional Planck units, of the cavity surrounding the BH. This supports our earlier statement about ``naturalness'' of the choice \eqref{DilaCharge}.

\subsection{Dilaton Chemical Potential and Surface Pressure}
\label{sec:DilatonChemPot}

The chemical potential associated with the dilaton charge \eqref{DilaCharge} is given by
\begin{equation}\label{pressure}
  \psi_c = - \left. \frac{\partial F_c}{\partial D_c} \right|_{T_c} ~.
\end{equation}
The variation of $F_c$ at fixed $T_c$ can be expressed as
\eq{
dF_c\Big|_{T_c} = -2\pi\,T_c\,dX_h+\frac{1}{2}\,e^{-Q_c}\left(\frac{d\killing_0}{\sqrt{\killing_0}}-\frac{d\killing_c}{\sqrt{\killing_c}}\right)-U_c e^{-Q_c}\left(\sqrt{\killing_0}-\sqrt{\killing_c}\right)dD_c\,.
}{eq:angelinajolie}
Using \eqref{dX_and_dT}, \eqref{dAc} and \eqref{dD_T} we find
\begin{equation}\label{eq:td7}
  \psi_c = - \frac{1}{2}\,U_c \, e^{-Q_c} \, \left(\sqrt{\killing_c} - \sqrt{\killing_0}\right)
   + \frac{1}{2} \, w_c'\, \left( \frac{1}{\sqrt{\killing_c}} - \frac{1}{\sqrt{\killing_0}}\right) ~.
\end{equation}
We have defined the dilaton chemical potential with a minus sign in \eqref{pressure} so that it follows the same conventions as a pressure in standard thermodynamics. This analogy becomes precise in several examples where $D_c$ plays the role of the area of a cavity surrounding the horizon.

The dilaton chemical potential \eqref{eq:td7} exhibits some interesting properties: It diverges with the Tolman factor as the horizon is approached. For the ground state solution $\mass=0$ it vanishes.
Finally, it transforms covariantly with the dilaton charge $\bar{D}_c(D_c)$, viz., $\bar{\psi}_c=\psi_c \,dD_c/d\bar{D}_c$. With the choice \eqref{eq:dc1} the dilaton chemical potential simplifies to
\eq{
\bar{\psi}_c = \eta\sqrt{\frac{\killing_0}{\killing_c}}+(1-\eta)\sqrt{\frac{\killing_c}{\killing_0}}-1
}{eq:dc2}
where
\eq{
\eta := \frac{1}{1-U_cw_c/w'_c}\,.
}{eq:dc3}
This definition is particularly useful if $\eta$ is constant, such as for models with $U=0$, Minkowskian ground state models and the so-called $ab$-family. We shall discuss them in detail in section \ref{sec:2DExamples}.

It is sometimes of interest to determine the values of the dilaton charge for which the dilaton chemical potential \eqref{eq:td7} vanishes. This can be thought of as extremizing the free energy with respect to the dilaton charge. The dilaton chemical potential vanishes if either of the following conditions is satisfied
\eq{
\sqrt{\killing_c}=\sqrt{\killing_0}\,,\qquad \sqrt{\killing_c}=-\frac{w'_c}{w_cU_c}\,\sqrt{\killing_0}
~.}{eq:td124}
The first possibility can only be realized asymptotically. Thus, one natural value of the cut-off is the asymptotic region, which might have been anticipated. The second possibility arises only if $-1<w'_c/(w_cU_c)<0$. In the latter case, the system is only stable against changes in $D_c$ if the free energy is a minimum. This is equivalent to positivity of the isothermal compressibility \eqref{eq:td110} discussed below.

\subsection{Internal Energy and First Law}
\label{sec:Energy}

All calculations so far have referred to the canonical ensemble, with the temperature $T_c$ and dilaton charge $D_c$ at the cavity wall held fixed. We can now perform a Legendre transformation to obtain the internal energy $E_c$ as a function of the entropy and the dilaton charge
\begin{equation}\label{Ec}
  E_c(S, D_c) = F_c(T_c, D_c) + T_c \, S ~.
\end{equation}
From \eqref{FreeEnergy} and \eqref{Entropy2} we obtain
\eq{
E_c = e^{-Q(D_c)}\left(\sqrt{\killing_0}-\sqrt{\killing_c}\right) \geq 0\,.
}{eq:td10}
The last inequality holds because we assume $\mass,w_c \geq 0$.
The internal energy $E_c$ agrees with the proper energy obtained from the quasi-local stress tensor of Brown and York \cite{Brown:1991gb}. 
This stress tensor is defined as
\begin{equation}
   T^{ab} \defeq \frac{2}{\sqrt{\gamma}} \, \frac{\delta \Gamma}{\delta \gamma_{ab}} ~.
\end{equation}
Using \eqref{dGamma} obtains
\begin{equation}
   T^{ab} = 2\,\left(\pi^{ab} - p^{ab}\right) ~.
\end{equation}
Contracting $T^{ab}$ with two copies of the unit-norm Killing vector $u_{a} = \delta_{a}^{\,\tau}$ gives the proper (Euclidean) energy density. This can be read off from \eqref{Momenta} and \eqref{eq:test}
\begin{equation}
   T^{ab} u_a u_b = e^{-Q(D_c)}\,\left(\sqrt{\xi_0} - \sqrt{\xi_c} \right) ~.
\end{equation}
This is precisely the internal energy given in \eqref{eq:td10}. Notice that it is not the same as the conserved charge associated with the Killing vector $\partial_{\tau}$,
\begin{equation}
   Q_{\partial_{\tau}} = \lim_{X_c \to \infty} \sqrt{\xi_c} \, T^{ab} u_a u_b = M\,,
\end{equation}
where $\sqrt{\xi}$ is the lapse function\footnote{Here 'lapse function' refers to the standard temporal evolution, to be distinguished from the lapse function of the radial evolution introduced in appendix \ref{sec:GaugeIndependence}.} in the $1+1$ decomposition of the metric \eqref{metric}.
The definition of the conserved charge $Q_{\partial \tau}$ involves the limit $X_c \to \infty$, which should be thought of as the two-dimensional analog of the limiting procedure used to integrate over a sphere at spatial infinity in higher dimensions. The asymptotic limit of the internal energy is proportional to the conserved charge $M$, rescaled by the Tolman factor
\eq{
\lim_{w_c \to \infty}E_c = \lim_{w_c \to \infty}\frac{\mass}{\sqrt{\killing_c}}\,.
}{eq:td15}
At finite values of $X_c$ the result \eqref{eq:td10} can be used to express the conserved quantity $M$ as a function of the internal energy
\eq{
 M = \sqrt{\xi_0}\,E_c - \frac{1}{2\,w_c}\,(\sqrt{\xi_0}\,E_c)^2 ~.
}{eq:td144a}
This has a simple interpretation in space-times where the Killing norm $\sqrt{\killing}$ approaches a constant, which we may set to unity without loss of generality. In that case $M$ coincides with the ADM mass, which is related to the internal energy in the region $X \leq X_c$ by
\eq{
M  = E_c - \frac{E^2_c}{2w_c}\,.
}{eq:td144}
The ADM mass contains two contributions: the first one is just the internal energy $E_c$ and the second one is the gravitational binding energy associated with collecting the energy $E_c$ in a region determined by $X_c$. Consistently, the second term is absent asymptotically.

Using \eqref{dAc} and the results for entropy \eqref{Entropy2} and dilaton chemical potential
\eqref{eq:td7} one can show that the internal energy obeys the first law of
thermodynamics~\footnote{A quasilocal form of the first law was obtained in
\cite{Creighton:1995uj}, using background subtraction to address the divergences in the action
and conserved quantities.}
 \eq{
\extd E_c = T_c \,\extd S - \psi_c \,\extd D_c ~. }{eq:td12} This quasilocal form of the first
law holds regardless of the particular model, the location $X_c$ of the cavity wall, or choice
of the dilaton charge. In this regard it contains a great deal more information about the
system than the microcanonical expression $dM = T dS$.

\subsection{Specific Heat at Constant Dilaton Charge and Fluctuations}
\label{sec:specificheat}

Specific heat at constant dilaton charge is given by
\eq{
C_D =\left. \frac{\partial E_c}{\partial T_c}\right|_{D_c} = \left.T_c \,\frac{\partial S}{\partial T_c}\right|_{D_c}\,.
}{eq:td18a}
Using \eqref{eq:td10}, \eqref{dX_and_dT} and \eqref{dA_D} we find
\eq{
C_D =  \frac{4\pi\, w_h' (w_c - 2 M)}{2 w_h'' (w_c - 2 M) + {w_h'}^2}\,.
}{eq:td18}
If $w''_h\neq 0$ then the asymptotic limit of $D_{D}$ reduces to the microcanonical specific heat
\eq{
\lim_{w_c\to\infty}C_D = 2\pi\, \frac{w'_h}{w''_h} = -\frac{\left(\frac{dS}{d\mass}\right)^2}{\frac{d^2S}{d\mass^2}}\,.
}{eq:td100}
We have implicitly assumed in the previous subsections that the specific heat is positive. This assumption will be examined in more detail in subsection \ref{sec:Stability}, but we can offer a partial justification by considering the near horizon limit of \eqref{eq:td18}. If the classical solution determined by \eqref{ClassicalSolutions} satisfies $w_c=w_h+\epsilon$, with $0<\epsilon\ll 1$, then the specific heat simplifies to
\eq{
C_D = \frac{\epsilon}{T}\,.
}{eq:td102}
Remarkably, it is always positive and vanishes as the cut-off approaches the horizon. This statement is model independent.

If the specific heat is positive and the temperature is large enough then one can take into account the effect of thermal fluctuations of the internal energy. The temperature is `large enough' if quantum fluctuations are considerably smaller than thermal fluctuations. Quantum fluctuations are typically of the order $\Delta_q E_c \approx 1$, whereas thermal fluctuations, $\Delta_t E_c= \sqrt{\langle E_c^2\rangle-\langle E_c\rangle^2} = \sqrt{C_D} T_c$, depend on specific heat at constant dilaton charge and on temperature. Thermal fluctuations dominate over quantum fluctuations if $C_D T^2_c\gg 1$. The canonical entropy, which takes into account thermal fluctuations, is given by the well-known result
\eq{
S_{\rm fluct} = S + \frac{1}{2} \ln{C_D T^2_c} + \dots
}{eq:td19}
Despite the fact that $T_c\to\infty$ at the horizon, the fluctuations remain finite there
\eq{
\lim_{w_c\to 2\mass}C_D T^2_c = e^{-Q_h} T\,.
}{eq:td104}
For some models (including the Schwarzschild BH) \eqref{eq:td104} does not satisfy $C_D T^2_c\gg 1$, even for large BH masses. In these cases thermal fluctuations do not dominate over quantum fluctuations near the horizon.

\subsection{Free Enthalpy and Gibbs-Duhem Violation}
\label{sec:Gibbs}

With the exception of the dilaton chemical potential, the thermodynamical quantities obtained in the previous subsections do not depend on the specific choice of the dilaton charge. This is not the case for quantities calculated in this and the following subsection. Quantities without a bar on top either refer to the dilaton charge \eqref{DilaCharge} or are independent from it; quantities with a bar refer to the dilaton charge \eqref{eq:dc1}.

We start with the free enthalpy (Gibbs function),
\eq{
G_c(T_c,\psi_c)=F(T_c,D_c)+\psi_cD_c=E_c(S,D_c)-T_cS+\psi_cD_c ~.
}{eq:td130}
Even though we have Legendre-transformed the internal energy with respect to the quantities that usually are considered extensive (entropy and ``volume'', viz., dilaton charge) the result is, in general, not zero. This can be thought of as a `violation' of the standard form of the Gibbs-Duhem relation~\footnote{For a discussion of alternative forms of this relation relevant for BHs, see \cite{Gibbons:2004ai}.}. There are, however, a few notable exceptions which we shall discuss in sections \ref{sec:2DExamples} and \ref{sec:StringExamples}. This shows that extensitivity properties are not necessarily as in standard thermodynamics. The physical reason underlying these observations is the existence of gravitational binding energy, as in \eqref{eq:td144}. Notably, free enthalpy is finite at the horizon, despite the fact that free energy diverges there:
\eq{
G_h = \left.\frac{1}{\sqrt{\killing_0}}\left(2M-TS+\frac{1}{2\pi}\,\mass SU_c \right)\right|_{D_c=X_h}\,.
}{eq:td132}
For comparison it is instructive to calculate
\eq{
\bar{G}_c = \frac{1}{\sqrt{\killing_c}}\left(2\eta\,M-TS\right)\,,
}{eq:dc5}
with $\eta$ as defined in \eqref{eq:dc3}. For constant $\eta$ the quantity $\sqrt{\killing_c} \bar{G}_c$ exclusively depends on properties of the horizon and the Gibbs-Duhem relation holds if $2\eta M=TS$.  The result \eqref{eq:dc5} is not only considerably simpler than the expression for $G_c$ (which we omitted to present) but it is also divergent at the horizon because of the Tolman factor. This is a clear demonstration of the fact that free enthalpy crucially depends on the choice of the dilaton charge.

The enthalpy
\eq{
\bar{\enthalpy}_c(S,\psi_c) = E_c(S,D_c)+\bar{\psi}_c\bar{D}_c =\bar{G}_c(T_c,\psi_c)+T_cS
}{eq:dc6a}
can be calculated from free enthalpy \eqref{eq:dc5} with the result
\eq{
\bar{\enthalpy}_c=2\eta\,\frac{\mass}{\sqrt{\killing_c}}\,.
}{eq:dc6}
It is positive if $\eta>0$ and also scales with the Tolman factor. The quantity $\enthalpy_c$ is much more complicated so we do not present it here.

\subsection{Miscellaneous Thermodynamical Quantities}
\label{sec:Misc}

The isothermal compressibility
\eq{
\isoth = -\frac{1}{D_c}\,\left.\frac{\partial D_c}{\partial\psi_c}\right|_{T_c}
}{eq:td110}
requires the calculation of $\partial\psi_c/\partial D_c|_{T_c}$ performed in appendix \ref{sec:Formulas}, cf.~\eqref{eq:td141}. Although the dilaton chemical potential diverges at the horizon isothermal compressibility in general remains finite there,
\eq{
\lim_{w_c\to 2\mass}\isoth = \frac{2\pi}{S}\, \cdot\,\frac{(2M)^{3/2}e^{Q_h/2}}{w''_hM-4\pi^2T^2-4\pi TMU_h-2M^2U'_h+M^2U^2_h} ~.
}{eq:td113}
We do not include the result for $\bar{\isoth}$ here, but it may be deduced from the results for $C_D$ and $\bar{C}_\psi$.

Specific heat at constant dilaton chemical potential is defined by
\eq{
C_\psi =\left.\frac{\partial \enthalpy_c}{\partial T_c}\right|_{\psi_c}=\left.T_c \frac{\partial S}{\partial T_c}\right|_{\psi_c}\,.
}{eq:td135}
Using \eqref{eq:useful} the difference between the specific heats,
\eq{
C_\psi-C_D = \isoth\,D_cT_c \left(\left.\frac{\partial \psi_c}{\partial T_c}\right|_{D_c}\right)^2\,,
}{eq:td136}
turns out to be positive if and only if isothermal compressibility is positive. The explicit calculation of $C_\psi$ is somewhat lengthy. A simpler result is obtained starting with the dilaton charge \eqref{eq:dc1}:
\eq{
\bar{C}_\psi = 2\pi\,\frac{w_h'}{w_h''}\,.
}{eq:td146}
This coincides with the asymptotic limit \eqref{eq:td100} and with the microcanonical specific heat. The difference between the specific heats is
\eq{
\bar{C}_\psi - C_D = 2\pi\,\frac{w_h'}{w_h''} \, \cdot\, \frac{{w_h'}^2}{{w_h'}^2+2w_h''(w_c-2\mass)}\,.
}{eq:td147}
For generic thermodynamical systems mechanical stability requires positive isothermal compressibility or, equivalently [cf.~\eqref{eq:td136}], $C_\psi>C_D>0$. However, as we have demonstrated the quantity $C_\psi$ depends crucially on the definition of the dilaton charge. Thus, unless there is a good physical reason to prefer one particular dilaton charge over all other possible choices, we cannot provide an unambiguous answer to the question of mechanical stability. It is intriguing that, for the dilaton charge \eqref{eq:dc1}, thermodynamic stability implies mechanical stability: $\bar{\kappa}_{T}>0$. This is because in a region where both $C_D$ and $\bar{C}_{\psi}$ are positive their difference $\bar{C}_\psi-C_D$ is also positive, as evident from \eqref{eq:td147}.

Additional thermodynamical quantities can be deduced from the ones we have presented here. One can basically exploit the results of standard thermodynamics by identifying the dilaton charge with ``volume'', and the dilaton chemical potential with ``pressure''. A few examples will be provided in section \ref{sec:2DExamples}. We conclude with a supplementary observation regarding an interesting geometric description of the thermodynamics of equilibrium systems due to Ruppeiner \cite{RevModPhys.67.605}. 
The main idea is to introduce a metric on the thermodynamic state space,
\eq{
ds^2_R := G_{IJ}^R dX^I dX^J\,,\qquad G_{IJ}=-\partial_I\partial_J S(X)\,,\qquad X^I=(E_c,D_c,\dots)\,,
}{eq:Rup1}
and to relate its geometric properties to thermodynamic quantities. The entropy $S$ here is expressed as a function of extensive variables $X^I$, which in our case comprise internal energy \eqref{eq:td10}, the dilaton charge \eqref{DilaCharge}, and any additional extensive quantities associated with generalizations of the metric-dilaton system.
This formalism was applied to various BH systems and led to some intriguing results \cite{Arcioni:2004ww}. 
It is worth mentioning that even in the absence of $U(1)$-charges the Ruppeiner metric is two-dimensional, which differs from the microcanonical analysis in \cite{Aman:2006mn}, where the Ruppeiner metric was two-dimensional for charged BHs only.

\subsection{Charged Black Holes}
\label{sec:Maxwell}

Gauge fields in two dimensions have no propagating physical degrees of freedom \cite{Birmingham:1991ty}, which makes the generalization of our results to charged BHs straightforward. We consider the addition of a single abelian gauge field, but the extension to several abelian or non-abelian gauge fields is straightforward. The improved action for dilaton gravity coupled to a gauge field is
\begin{align}\label{actionmax}
  \Gamma = &- \frac{1}{2} \int_{\MM} \bns d^{2}x \sqrt{g} \, \left[ X\,R - U(X) \, (\nabla X)^2 - 2 \,V(X)\right]
+ \int_{\MM} \bns d^{2}x \sqrt{g} \, f(X)\,F_{\mu\nu}F^{\mu\nu} \nonumber \\
       & - \int_{\dM} \bns dx \sqrt{\ga} \, X \, K
        + \int_{\dM} \bns dx \sqrt{\ga} \, e^{-Q(X)}\sqrt{\killing_0(X)}~.
\end{align}
The abelian field strength $F_{\mu\nu}=\nabla_\mu A_{\nu}-\nabla_\nu A_{\mu}$ is coupled to the dilaton field via the function $f(X)$. If this function is constant then we say that the gauge field is minimally coupled, and non-minimally coupled otherwise. We emphasize that the gauge field $A_\mu$ cannot contribute to the boundary counterterm because we require gauge- and diffeomorphism-invariance. The only gauge-invariant scalar that is linear in $A_\mu$ or its derivatives comes from the contraction of the field strength with the $\epsilon$-tensor: $F_{\mu\nu}\epsilon^{\mu\nu}$. However, this term contains a derivative normal to the boundary and is not allowed in a counter-term of the form \eqref{CT1}. All other gauge-invariant scalars constructed from $A_\mu$, like $F_{\mu\nu}F^{\mu\nu}$, necessarily contain such derivatives and are also ruled out.

The solution of the Maxwell equation $\nabla_{\mu}(f(X)\,F^{\mu\nu})=0$ is
\eq{
F^{\mu\nu}=\frac{q}{4 f(X)}\,\epsilon^{\mu\nu} \, ,
}{eq:max2}
where $q$ is the electric charge, and the factor of $\tfrac{1}{4}$ has been introduced for convenience. When working in the diagonal gauge \eqref{metric} we use the convention $\epsilon^{\tau r}=-\epsilon^{r\tau}=+1$. The equations of motion for the metric and dilaton can be solved in a manner similar to uncharged BHs. The metric and dilaton are again given by \eqref{metric} and \eqref{XPrimeDef}, but the square of the Killing norm is now
\eq{
\killing(X)=e^{Q(X)}\left(w(X)-2\mass + \frac{1}{4}\,q^2\,h(X)\right)
}{eq:max1}
The `ground state' is defined as the solution with vanishing mass and charge, $\mass=q=0$, leading to the same Killing norm $\sqrt{\killing_0}$ as in \eqref{GroundState}. The function $h(X)$ is defined as
\eq{
h(X):=\int^X \bns d\tilde{X}\,\frac{e^{Q(\tilde{X})}}{f(\tilde{X})}\,,
}{eq:max17}
where the integral is evaluated at $X$ and the integration constant is absorbed into the definition of $w$. The result \eqref{eq:max1} could also have been obtained from our earlier solution for the uncharged BH, by replacing $V(X)$ in \eqref{MetricEOM} and \eqref{XEOM} with an effective potential of the form
\eq{
V^{\rm eff}(X) = V(X) - \frac{q^2}{8f(X)}\,.
}{eq:max3}
We shall return to this point at the end of this subsection.

To obtain the on-shell action we use
\eq{
\int_{\MM} \bns d^{2}x \sqrt{g} \,fF^{\mu\nu}F_{\mu\nu} = 2\int_{\MM} \bns d^{2}x \sqrt{g} \,\nabla_{\mu}(fF^{\mu\nu}A_\nu)-2\int_{\MM} \bns d^{2}x \sqrt{g} \,A_\nu\nabla_{\mu}(fF^{\mu\nu})
}{eq:max4}
and note that the second term vanishes on-shell. The first term, a total derivative, can also be interpreted as one-dimensional Chern-Simons terms with support at both the boundary and the horizon
\begin{align}
   2 \int_{\MM}\bns d^{\,2}x \sqrt{g} \, \nabla_\mu (f F^{\mu\nu}A_{\nu})
   & = -\frac{q}{2}\, \int \nts \left. A_\tau d\tau \raisebox{14pt}{}\right|^{X_c}_{X_h}
     = - \frac{q}{2} \, \left( CS_1(X_c) - CS_1(X_h)\raisebox{11pt}{}\right) \nonumber \\
   & = -\frac{q}{2} \, \beta \, \left(A_\tau(X_c)-A_\tau(X_h)\raisebox{11pt}{}\right)\,. \label{eq:CS1}
\end{align}
It is convenient to fix the gauge so that $A_{r}=0$. In that case the $\tau$ component of the gauge field is given by
\eq{
A_\tau (X) =-\frac{q}{4} \, \left(h(X) - h(X_h) \right) + A_{\tau}(X_h)\,.
}{eq:maxgauge}
This is just the potential difference between the cavity wall and the horizon, plus an arbitrary constant. The proper electrostatic potential between the cavity wall and the horizon is related to $A_{\tau}$ by
\eq{
\Phi(X) := \frac{A_{\tau}(X)-A_\tau(X_h)}{\sqrt{\killing(X)}} ~.
}{eq:max6}
Note that it contains a Tolman factor, like the proper temperature.

We are now ready to present the generalization of \eqref{CutOffAction} to the case of charged BHs,
\eq{
\Gamma_c = \beta_c \left(e^{-Q_c}\left(\sqrt{\killing_0}-\sqrt{\killing_c}\right)-2\pi X_h T_c -q \Phi_c\right) ~.
}{eq:max5}
As pointed out in \cite{Davis:2004xi} the logarithm of the partition function,
\eq{
Y_c(T_c,D_c,\Phi_c)= T_c\, \Gamma_c = - 2\pi\,X_h \, T_c + e^{-Q_c}\,\left(\sqrt{\killing_0}-\sqrt{\killing_c}\right) - q \, \Phi_c\,,
}{eq:Y}
is not the Helmholtz free energy. Rather, $Y_c$ is the Legendre transform of $F_c$ with respect to the proper electrostatic potential $\Phi_c$. This is because the Maxwell equation is obtained by varying the action with respect to the gauge field, so it is the gauge field (and not the field strength) that obeys Dirichlet conditions at $\dM$. The on-shell action is therefore a function of $A_{\mu}$ evaluated at the boundary. The derivative of $Y_c$ with respect to $\Phi_c$, holding $T_c$ and $D_c$ fixed, gives the conserved electric charge
\eq{
 \left. -\frac{\partial Y_{c}}{\partial \Phi_c} \right|_{D_c,T_c} = q ~.
}{eq:ElectricCharge}
The Legendre transformation
\eq{
 F_c(T_c,D_c,q) =  Y_c(T_c,D_c,\Phi_c) + q \, \Phi_c
}{eq:max7}
leads to the same expressions for the Helmholtz free energy \eqref{FreeEnergy} and consequently also for entropy \eqref{Entropy2} and internal energy \eqref{eq:td10}.

For minimally coupled Maxwell fields the electrostatic potential $\Phi \propto (X_c-X_h)$ is proportional to the separation between the cavity wall and the horizon. If the Killing norm asymptotes to unity then the proper electrostatic potential $\Phi_c$ diverges in the $X_c \to \infty$ limit.
As a result, the improved action for these models diverges in the limit where the cavity wall is removed from the system. We emphasize that this divergence reflects the physics of the linear electrostatic potential and therefore cannot be removed from the action. The same rule that was mentioned at the end of subsection \ref{sec:BlackHoles} applies here, as well. One does not obtain the correct improved action by simply dropping divergent terms. In the present case, the action depends on the potential difference between the wall and the horizon, which is gauge invariant. Dropping the divergent term from the action leads to a violation of gauge invariance.
An example of a charged BH with minimal coupling is provided by 2D type 0A string theory \cite{Davis:2004xi}, where $e^{-Q}=w=\lambda X$ and $f=\pi\alpha^\prime$. As an example of non-minimal coupling we mention the Reissner-Nordstr\"om BH, $e^{-Q}=w=2\sqrt{X}$ and $f(X)=X$ (with convenient normalizations for $e^Q$ and $f$). Our formula \eqref{eq:max17} gives $h=-1/\sqrt{X}=-1/r$ in this case, which is the correct Coulomb-potential. Thus, the typical linear (confining) behavior of the electrostatic potential in two dimensions can be modified by a non-minimal coupling $f(X)$.

Maxwell fields open up an interesting possibility. Consider a theory without Maxwell fields whose potential $V$ takes the form
\eq{
V(X)=\VV(X) - \frac{q^2}{8f(X)} ~.
}{eq:max10}
We can also think of this potential as the result of integrating out a gauge field with coupling function $f(X)$, in a theory with dilaton potential $\VV(X)$. We simply re-interpret \eqref{eq:max10} as \eqref{eq:max3}. These two points of view will generally lead to different conclusions about the thermodynamics. In particular, the ``natural'' choice of ground state is different in the two formulations. If we treat the system as dilaton gravity with a Maxwell field then the natural ground state is given by $M=q=0$. But in the original description, with no Maxwell fields, the ground state is associated with $M=0$. For instance, the ground state $M=q=0$ is Minkowski space for the Reissner-Nordstr\"om BH, whereas in the alternative description without Maxwell field the natural ground state $\mass=0$ is a BPS solution. It is straightforward to generalize these considerations to potentials of the form
\eq{
V = \sum_{n=-\infty}^{\infty} \lambda_n X^{n/a}\,,
}{eq:max8}
where $a$ is some natural number. This applies to almost all models in the literature. Now we may introduce Maxwell fields to convert all but one of the terms in the generalized Laurent series into Maxwell terms. This leads to a rather simple thermodynamical description and may be useful in various contexts. We shall provide an example in \ref{sec:BTZ} when we consider the toroidal reduction of the BTZ BH.

\subsection{Conformal Properties and Microcanonical Thermodynamics}
\label{sec:Conformal}

In appendix \ref{sec:NewVariables} we discuss some consequences of dilaton dependent Weyl rescalings \eqref{newmetric}. If interpreted as a conformal transformation,\footnote{We use the term `conformal transformation' in this section, as commonly done in the literature, though \eqref{newmetric} should really be referred to as a Weyl transformation.} rather than as a change of variables, a new action \eqref{Action} is obtained with potentials given by \eqref{eq:apB7}. We say that these models are conformally related to each other and sometimes refer to them as representing different `conformal frames' within an equivalence class of models. We consider now the transformation behavior of thermodynamical quantities. To this end it is very helpful to note \eqref{eq:apB13} that $w$ is conformally invariant. Therefore, quantities which depend only on $w$ and its derivatives are conformally invariant, but quantities that depend on $Q$ are not.

The conformally invariant thermodynamical quantities comprise the on-shell action $\Gamma_c$ \eqref{CutOffAction}, surface gravity $T$ \eqref{T}, mass parameter $\mass$ \eqref{Horizon} and entropy $S$ \eqref{Entropy2}. It is interesting to note that the specific heat at constant dilaton charge $C_D$ \eqref{eq:td18} is conformally invariant, despite the fact that both $E_c$ and $T_c$ depend on the conformal frame. This is because the temperature $T_c$ \eqref{Tc}, free energy $F_c$ \eqref{FreeEnergy} and internal energy $E_c$ \eqref{eq:td10} are conformally covariant, with a conformal weight of $-1$ due to the Tolman factor $1/\sqrt{\killing_c}$. With this convention the metric is defined to have weight $+2$. Quantities that are neither invariant nor covariant are the dilaton chemical potential $\bar{\psi}_c$ \eqref{eq:dc2}, free enthalpy $\bar{G}_c$ \eqref{eq:dc5}, enthalpy $\bar{\enthalpy}_c$ \eqref{eq:dc6} and isothermal compressibility $\isoth$ \eqref{eq:td110}. These statements are independent from the choice of the dilaton charge. An interesting issue arises for the specific heat at constant dilaton chemical potential: while the result for $\bar{C}_\psi$ \eqref{eq:td146} is conformally invariant, this is not the case for $C_\psi$ with a generic choice of dilaton charge $D_c$. This provides another rationale for working with the charge $\bar{D}_c$, as defined in \eqref{eq:dc1}.

As a simplification one can consider microcanonical thermodynamics (or ``horizon thermodynamics'') without referring to an actual observer \cite{Gegenberg:1994pv}. If one is interested only in the microcanonical entropy $S$, surface gravity $T$, and their relation to the mass parameter $\mass$, then one can use a conformal transformation to set $U=0$. In this scenario the two relevant formulas, besides the result for entropy \eqref{Entropy2}, are the Smarr formula for two-dimensional dilaton gravity,
\eq{
M = \frac{w(S/2\pi)}{2}
}{eq:Smarr}
and the microcanonical first law
\eq{
dM = T dS\,.
}{eq:1stlaw}
It is important to realize that the microcanonical thermodynamics is not equivalent to canonical thermodynamics in a limit where the dilaton charge approaches some specific value, with rare exceptions. The two pictures are equivalent only if there is a value of the dilaton charge where the dilaton chemical potential vanishes, $\psi_c=0$, the proper temperature coincides the with Hawking temperature, $T=T_c$, and the canonical ensemble is well-defined. In that case the canonical first law \eqref{eq:td12} coincides with the microcanonical one \eqref{eq:1stlaw}, and the mass $\mass$ coincides with the internal energy $E_c$. The first two conditions are satisfied in the asymptotic region of models with a Minkowskian ground state. But the third condition is typically violated: in these models the canonical ensemble quite often is not defined in the $X_c \to \infty$ limit. We show now why this is the case and under which conditions the cut-off can be removed.

\subsection{Stability and Phase Transitions}
\label{sec:Stability}

The most important issue related to stability in the canonical ensemble is the effect of thermal fluctuations. If the specific heat at constant dilaton charge is negative, $C_D < 0$, then thermal fluctuations in the internal energy have an imaginary component and the system is unstable. From the point of view of the path integral, $C_{D} < 0$ implies that the Gaussian integral around a particular extremum of the action does not converge. Fortunately, the assumptions $w_c > w_h$ and $w_h'>0$ guarantee that there is an open region around the horizon where the specific heat \eqref{eq:td102} is positive. This result is model-independent. Thus, if the horizon is located sufficiently close to the cavity wall then the canonical ensemble is always well-defined. Of course, a quantitative definition of `sufficiently close' will depend on the features of the particular model.

In most models a classical solution is only stable when the cavity wall is located at an $X_c$ that is less than some finite, critical value of the dilaton. Consider a BH associated with a solution $M$ of the condition \eqref{ClassicalSolutions}. We assume that this system has a positive specific heat at constant dilaton charge. Now we want to change the boundary conditions $X_c$ and $\beta_c$ while holding $M$ fixed. As described at the beginning of section \ref{sec:Thermo}, this is accomplished by moving $X_c$ away from the horizon while simultaneously relaxing $\beta_c$ according to \eqref{ClassicalSolutions}. The equation \eqref{eq:td18} allows us to make a quantitative statement about the point at which this process destabilizes the system. Suppose that $w$ is a monotonic function. Then a solution $X_{\rm crit}>X_h$ of
\eq{
w(X_{\rm crit}) = 2M - \frac{{w_h'}^2}{2w_h''}
}{eq:stab1}
represents a critical value of the dilaton where the specific heat diverges and changes sign. A well-defined canonical ensemble containing the classical solution exists for any value $X_h < X_c < X_{\rm crit}$, but not for $X_c > X_{\rm crit}$. A similar argument can be made for $w$ that are not monotonic. Notice that in either case the specific heat approaches zero at $X_c = X_h$. This reflects the fact that the system is only defined for a cavity wall that is external to the horizon.

In general, the condition \eqref{ClassicalSolutions} identifies multiple classical solutions consistent with the boundary conditions \cite{York:1986it}. We refer to each solution as a `branch' of \eqref{ClassicalSolutions}, denoted by $M_{i}(X_c,\beta_c)$. The thermodynamic stability of each branch can be determined by looking for a solution $X_{\rm crit}^{i} > X_h$ of \eqref{eq:stab1}. There are three possibilities:
\begin{enumerate}
\item There is no solution $X_{\rm crit}^{i} > X_h$. This implies that $C_{D}$ is positive, and the branch is thermodynamically stable.
\item There is a solution $X_{\rm crit}^{i} > X_c > X_h$. Again, the specific heat is positive and the branch is thermodynamically stable.
\item There is a solution such that $X_c > X_{\rm crit}^{i} > X_h$. In this case the specific heat is negative, and the branch is not thermodynamically stable.
\end{enumerate}
In the third case the solution $M_{i}(X_c, \beta_c)$ may decay into another field configuration with lower free energy via a Hawking-Page phase transition \cite{Hawking:1982dh}. Because these considerations involve only $w$ the emergence of a phase transition happens in any conformal frame.

We are now able to address the issue of the $X_c \to \infty$ limit in more detail. Given a branch $M_{i}$ of \eqref{ClassicalSolutions} that is thermodynamically stable for some initial values of the boundary conditions, we are interested in the process described by
\begin{equation}\label{Process}
   X_c \to \infty \quad \quad \beta_c \to \lim_{X_c \to \infty} \sqrt{\xi(D_c,M_{i})}\,\beta(M_{i})
\end{equation}
while holding $M_{i}$ fixed. This removes the cavity wall, leaving an isolated BH. The process is only defined if $\beta_c$ remains real and non-negative for all values of $X_c$. Furthermore, it may be defined on some branches but not others. Finally, even if the process \eqref{Process} is defined its endpoint may not be thermodynamically stable. The existence of a thermodynamically stable limit \eqref{Process} is far from being a generic feature in the models we consider. For instance, the Schwarzschild BH has $\lim_{X_c\to\infty} C_{D} < 0$. In models where the endpoint might be stable, the equation \eqref{eq:stab1} can be used to group solutions into two sectors. If \eqref{eq:stab1} admits a finite solution $X_{\rm crit} > X_h$ for a particular value of $\mass$, then that solution is unstable. But if there is no $X_{\rm crit} > X_h$ for that value of $\mass$ then the solution is stable. In higher dimensions the best known example of this phenomenon is the Schwarzschild-AdS solution, where the presence of a cosmological constant stabilizes very massive BHs without the need for an external reservoir. An analogous two-dimensional dilaton gravity model has $w_{SAdS}=2\sqrt{X}(1+X/\ell^2)$ with $X=r^2$. For large enough masses, such that $X_h>\ell^2/3$, the quantity $w''_h$ is positive and the system is stable, while for $X_h<\ell^2/3$ the quantity $w''_h$ becomes negative and the system is unstable. This can be translated into a critical temperature, where the specific heat diverges. It signals a phase transition \cite{Hawking:1982dh} between a BH phase for large masses (small temperatures) and a thermal AdS phase for small masses (large temperatures). We shall encounter similar examples in subsection \ref{sec:AAdS}.

\subsection{Tunneling and Constant Dilaton Vacua}
\label{sec:CDV}

In addition to the generic family of solutions \eqref{metric}, certain models may admit isolated solutions known as constant dilaton vacua (CDV). Namely, for each solution $X=X_0$ of $V(X)=0$ there is a CDV with a line element \eqref{metric} whose Killing norm $\sqrt{\killing}$ is given by
\eq{
\killing=c+ar-\frac12 \lambda r^2\,.
}{eq:CDV1}
Here $c$ and $a$ are constants of motion. The constant $\lambda$ is determined by
\eq{
R^{CDV}=-\partial_r^2\killing=\lambda=2V'(X_0)\,.
}{eq:CDV2}
Because they have a different number of constants of motion CDVs and the generic solutions belong to different 'superselection sectors'. Therefore there is no perturbative way in which a generic solution could decay into a CDV or vice versa. This is consistent with the result \cite{Bergamin:2003mh} that fully supersymmetric solutions necessarily are CDVs, while half BPS solutions can never be CDVs. If a supersymmetric CDV exists then the natural ground state solution in the generic sector is a half BPS solution, and both kinds of solutions are stable. However, once matter is switched on there could be non-perturbative interactions which allow tunneling from one sector into the other. We would like to determine whether such a tunneling between a generic solution and a CDV is possible, and if so, which of the two configurations is favored.

We immediately encounter an obstacle: CDVs are isolated in the sense that for an open interval of boundary conditions, $X_c\in(X_<,X_>)$, there is only a countable number of values of the dilaton consistent with the equations of motion. Typically this number is either zero or of order unity. Thus, at first glance CDVs are irrelevant because they seem to be a set of measure zero in the partition function \eqref{PartitionFunction}. Only for infinite finetuning, $X_c=X_0$, there may be a contribution from CDVs. In order to decide whether or not CDVs may be disregarded we consider an open interval of boundary conditions for the dilaton and the metric. This interval may be arbitrarily small, but it should include exactly one value $X_c=X_0$ that leads to a CDV.\footnote{In fact, in any experimental situation this smearing of boundary data is physically more reasonable than an infinite finetuning.} In order to get the number of classical solutions determined by this set of boundary data we take the product of the subset of initial conditions consistent with the classical equations of motion times the space of continuous constants of motion that may be adjusted freely. We call the dimension of this space ``weight'' and want to compare the weight of generic solutions with the weight of CDVs.
Let us work in fixed physical units. Then, for generic solutions \eqref{metric} there is only a discrete set of values for $\mass$ consistent with the boundary conditions on $X_c$ and $\killing_c$. So the weight is 2 (from the set of boundary conditions consistent with the equations of motion) plus 0 (from the space of continuous constants of motion that may be adjusted freely) and thus equals to 2. For CDVs the space of allowed boundary conditions is just one-dimensional since the dilaton is fixed to a certain value, but now there are two constants of motion, $c$ and $a$. Therefore, the space of constants of motion that may be adjusted freely is one-dimensional. We conclude that also for CDVs the weight is 2 and they should not be regarded as a set of measure zero. Below we use again sharp boundary conditions and assume $X_c=X_0$ for simplicity.

Now that we have convinced ourselves that CDVs are not a set of measure zero we address the question whether tunneling from a generic solution into a CDV is a favorable process or not. To this end we compare the generic result for the 'improved' on-shell action \eqref{CutOffAction} with the on-shell action for a CDV,
\eq{
\Gamma^{CDV}=-2\pi X_0\,\chi + \be\sqrt{\killing} e^{-Q(X_0)/2}\sqrt{w(X_0)}\,.
}{eq:CDV3}
The first term in \eqref{eq:CDV3} contains the Euler-characteristic $\chi$ of the manifold $\MM$. The last term is the universal counterterm \eqref{ICT} evaluated at $X=X_0$. It has the same structure and, for the same set of boundary data, also the same magnitude as the corresponding term in the 'improved' on-shell action for a generic solution. The periodicity in Euclidean time $\be$ is well-defined and positive provided the inequality $a^2+2c\lambda>0$ holds, which we assume henceforth. Minkowski or AdS vacua require a non-vanishing Rindler acceleration $a$ while de Sitter solutions ($c\lambda>0$) are compatible with $a=0$. We conclude that tunneling from a generic solution into a CDV is favored if the inequality
\eq{
X_0 \,\chi - X_h > \frac{2w(X_0)-2w(X_h)}{w'_h}
}{eq:CDV5}
holds. Here $X_h<X_0$ is the dilaton evaluated at the horizon of the generic solution. Since the right hand side is non-negative the Euler characteristic must be positive for this inequality to hold. Therefore we restrict the remaining discussion to the topologies of either a sphere ($\chi=2$) or a disk ($\chi=1$). The latter case is consistent with a BH interpretation, while the former case may arise for de Sitter. As an example for a model with enhanced tunneling to its CDV solution we mention $w(X)=c^3+(X-c)^3$. The CDV solution solution for this model is $X_0=c>0$, and the inequality \eqref{eq:CDV5} holds for any value of $\mass$ consistent with $X_0>X_h$. For generic models there are two interesting limits. If $X_0$ is close to the horizon, $X_0=X_h+\eps$, then for $\chi=2$ the inequality \eqref{eq:CDV5} holds and tunneling to a CDV is enhanced, whereas for $\chi=1$ tunneling is suppressed. If $X_0\gg 1$ then \eqref{eq:CDV5} simplifies to $X_0\chi > 2w(X_0)/w_h' + \dots$ Since we demand that $w\to\infty$ as $X\to\infty$ this inequality establishes a dichotomy in the space of dilaton gravity theories: if the function $w$ diverges asymptotically faster (slower) than linearly tunneling into a CDV is suppressed (enhanced), regardless of whether $\chi=1$ or $\chi=2$.

In summary, a transition between generic solutions and CDVs is possible only through tunneling. If the inequality \eqref{eq:CDV5} holds tunneling from a generic solution into a CDV is enhanced, otherwise it is suppressed. We remark that such a tunneling is far from being a generic feature: first of all, many models do not exhibit CDVs, and second, many of the models that do allow for CDV solutions violate the inequality \eqref{eq:CDV5} for all possible values of $X_h<X_0$. In fact, none of the models that we are going to discuss explicitly in sections \ref{sec:abFamily}, \ref{sec:StringExamples} and \ref{sec:HigherDimensionalExamples} allows for an enhanced tunneling into a CDV.

\section{Two Dimensional Examples}
\label{sec:2DExamples}

\subsection{Minkowski Ground State Models}
\label{sec:MGS}

If the relation
\eq{
e^Qw=1
}{eq:MGS1}
holds then $\xi_0 = 1$ and the ground state solution \eqref{GroundState} is (upon Wick rotation back to Lorentzian signature) Minkowski space-time. These models are of particular interest as they contain asymptotically flat solutions like the Schwarzschild and Witten BHs. As we shall see \eqref{eq:MGS1} leads to considerable simplifications.

The free energy
\eq{
 F^{MGS}_c =  E_c^{MGS} - \frac{S T}{\sqrt{1-\frac{2M}{w_c}}}
}{eq:td11}
and internal energy
\eq{
E_c^{MGS}=w_c\left(1-\sqrt{1-\frac{2M}{w_c}}\right)
}{eq:td13}
have a finite limit if the cut-off is removed, $\lim_{w_c \to \infty}F_c^{MGS} = M-TS$ and $\lim_{w_c \to \infty}E_c^{MGS} = M$. The asymptotic version of the first law \eqref{eq:td12} simplifies to the microcanonical $dM=TdS$. The chemical potential conjugate to the dilaton charge $\bar{D}_c^{MGS}=w_c$ \eqref{eq:dc1},
\eq{
\bar{\psi}_c^{MGS} = \frac12 \,\left(1-\frac{2M}{w_c}\right)^{1/2}+\frac12\, \left(1-\frac{2M}{w_c}\right)^{-1/2}-1 \geq 0~,
}{eq:dc2MGS}
is always non-negative and vanishes for the ground state solution only. It is useful to notice that the quantity defined in \eqref{eq:dc3} becomes a universal constant, $\eta=1/2$. For the formulation based upon the dilaton charge \eqref{DilaCharge} we get $\psi_c^{MGS} = \bar{\psi}_c^{MGS} w'_c$, which is also non-negative provided $w_c'\geq 0$. Asymptotically \eqref{eq:dc2MGS} implies $\lim_{w_c \to \infty}\bar{\psi}_c^{MGS} = M^2/(2w_c^2)$.
From \eqref{eq:dc5} we obtain the free enthalpy
\eq{
\bar{G}_c^{MGS} = \frac{M-TS}{\sqrt{1-\frac{2M}{w_c}}}\,.
}{eq:dc5MGS}
The relevant quantities for isothermal compressibility simplify to $\iscoa_c^{MGS}=w''_c$, $\iscob_c^{MGS}=-w''_c/2-w'_cN_c/2$, $\iscoc_c^{MGS}=-w''_c/2+w'_cN_c/2$ and $N_c=(2Mw'_cw''_h)/[w_c({w'_h}^2+2w''_h(w_c-2M))]$. Evaluation at the horizon yields $\lim_{w_c\to 2\mass}\isoth^{MGS} = 2\pi/(S w''_h)$.
In the alternative formulation the result is
\eq{
\bar{\isoth}^{MGS}=w_c\sqrt{1-\frac{2M}{w_c}}\,\cdot\,\frac{2w_c-4M+{w_h'}^2/w_h''}{4M^2}\,.
}{eq:dc7}
The zeros $X_c>X_h$ of $\bar{\isoth}^{MGS}$ coincide with the poles of $C_D^{MGS}$ \eqref{eq:td18}. We do not investigate these models in more detail here but shall encounter them again in this and the next section.

\subsection{Models in the $a$-$b$ Family}
\label{sec:abFamily}

The so-called $ab$-family of models \cite{Katanaev:1997ni}, \eq{ U=-\frac aX\,,\qquad V= -
\frac B2 X^{a+b}\,. }{eq:ab1} includes the dimensional reduction of spherically symmetric BHs
in $d+1$ space-time dimensions [$a=(d-2)/(d-1),b=-1/(d-1)$], the Witten BH ($a=1, b=0$)
\cite{Witten:1991yr}, the Jackiw-Teitelboim (JT) model ($a=0, b=1$) \cite{JT, Lemos:1996bq},
and many others. In particular, the Schwarzschild BH arises for $a=-b=1/2$. Minkowskian ground
state models obey the relation $a=1+b$. Models with $a=1-b$ have an (A)dS ground state and
models with $b=0$ a Rindler ground state. Even more general models often are approximated quite
well by the monomial potentials of the $ab$ family. Therefore our discussion here will reveal
some generic features of the thermodynamics of dilaton gravity in two dimensions. For the sake
of definiteness, and to comply with our working assumption \eqref{wAsymptotic}, we assume
$b>-1$ and $B>0$. In addition, we reduce clutter by imposing $B=b+1$, without loss of
generality. We fix the constants in \eqref{QDef} and \eqref{wDef} so that \eq{
w=X^{b+1}\,,\qquad e^Q=X^{-a}\,. }{eq:ab1.5} With these assumptions the BH horizon is located
at \eq{ X_h = (2\mass)^{1/(b+1)} ~. }{eq:ab2} Note that a negative value of $\mass$ would
result in a naked singularity at $X=0$ if $a<2$. Surface gravity yields \eq{ T =
\frac{b+1}{4\pi} (2\mass)^{b/(b+1)}\,, }{eq:ab3} which leads to a mass-to-temperature relation
$\mass\propto T^{1+1/b}$. The Killing norm squared and asymptotic Killing norm squared are
given by \eq{ \killing_c=\killing_0-2\mass X^{-a}_c\,,\qquad \killing_0 = X^{b-a+1}_c \geq
\killing_c\,. }{eq:ab4} If we demand $\xi_0\neq 0$ for $X_c\to\infty$ the inequality $b-a+1\geq
0$ must be met.

\subsubsection{Thermodynamics of the $a$-$b$ Family}

Entropy \eqref{Entropy2} reads
\eq{
S=2\pi (2\mass)^{1/(b+1)}\,.
}{eq:ab5}
The free energy \eqref{FreeEnergy} simplifies to
\eq{
F_c=X^a_c(\sqrt{\killing_0}-\sqrt{\killing_c})-\frac{(b+1)\mass}{\sqrt{\killing_c}}\,.
}{eq:ab6}
For $b<0$ free energy changes its sign at a particular value of the cut-off,
\eq{
\sqrt{\killing_c} = \frac{1+b}{1-b} \sqrt{\killing_0}\,,
}{eq:ab8}
and the asymptotic free energy is positive. This equation always has exactly one solution under the assumptions $-1<b<0$ and $\mass > 0$. For Schwarzschild this happens at $\sqrt{X}=9\mass/4$. With the dilaton charge $\bar{D}_c=X_c^{(a+b+1)/2}$ the dilaton chemical potential \eqref{eq:dc2} simplifies to
\eq{
\bar{\psi}_c=\frac{b+1}{a+b+1}\,\sqrt{\frac{\killing_0}{\killing_c}}+\frac{a}{a+b+1}\,\sqrt{\frac{\killing_c}{\killing_0}}-1\,.
}{eq:ab20}
It vanishes at a particular value of the dilaton given by
\eq{
\sqrt{\killing_c}=\frac{1+b}{a}\sqrt{\killing_0}\,.
}{eq:ab10}
This equation has a solution if $0<b+1<a$. For Minkowskian ground state models the dilaton chemical potential vanishes for $X_c\to\infty$ only. For (A)dS ground state models the dilaton chemical potential vanishes precisely at the value of the cut-off that leads to vanishing free energy. Internal energy \eqref{eq:td10} is given by
\eq{
E_c=X^{(a+b+1)/2}_c\left(1-\sqrt{1-2\mass X^{-1-b}_c}\right)\,.
}{eq:ab11}
The specific heat at constant dilaton charge \eqref{eq:td18} simplifies to
\eq{
C_D=S \,\frac{X^{b+1}_c-2\mass}{b X^{b+1}_c-(b-1) \mass}\,.
}{eq:ab13}
It changes sign at $X_{\rm crit}$ if the inequality
\eq{
X_{\rm crit}=\left(\mass\frac{b-1}{b}\right)^{1/(1+b)}>X_h
}{eq:ab14}
holds, which is possible for $-1<b<0$ only. Thermal fluctuations near the horizon are determined by
\eq{
\lim_{w_c\to 2M}C_DT_c^2 \propto M^{(a+b)/(b+1)}\,,
}{eq:MGS1783}
with a proportionality constant of order of unity. If $b>-a$ ($b<-a$) thermal (quantum) fluctuations dominate for large masses. If $a+b=0$ both kinds of fluctuations are relevant. The only Minkowskian ground state model with this property is the Schwarzschild BH. Specific heat at constant dilaton chemical potential \eqref{eq:td146},
\eq{
\bar{C}_\psi = \frac{S}{b}\,,
}{eq:MGS1784}
is positive for $b>0$ and coincides with the $X_c\to\infty$ limit of \eqref{eq:ab13}. The free enthalpy
\eq{
\bar{G}_c = (1-a-b)\,\frac{b+1}{a+b+1}\,\cdot\,\frac{\mass}{\sqrt{\killing_c}}
}{eq:td148MGS}
obeys the Gibbs-Duehm relation $\bar{G}_c=0$ for (A)dS ground state models only.
To get the isothermal compressibility \eqref{eq:td110} we calculate \eqref{eq:td111}:
$\iscoa_c = \frac 14 (a+b+1)(a+b-1)X^{(a+b-3)/2}_c$, $\iscob_c = \frac a2 X^{a-2}_c(1-a-X_cN_c)$, and $\iscoc_c = \frac 12 (b+1) X_c^{b-1}(-b+X_cN_c)$, with
\eq{
N_c=-\frac{b}{2X_c}\,\frac{2a\mass-(a-b-1)X^{1+b}_c}{(b-1)\mass-bX^{1+b}_c}\,.
}{eq:ab15}
For any model without (A)dS ground state ($a\neq 1-b$) the poles of $N_c$ and $C_D$ coincide with each other. For (A)dS ground state models $N_c=b/X_c$ and therefore $\isoth$ vanishes.

For Minkowskian ground state models we can discuss the life-time of a BH without specifying explicitly the degrees of freedom that carry the Hawking quanta. Let us assume that the two-dimensional Stefan-Boltzmann law, asymptotic flux$\,\propto T^2$, is valid. Then the mass loss per time is determined by $d\mass/dt\propto - T^2$. The proportionality constant depends on the matter content and on the physical units. With \eqref{eq:ab3} it is straightforward to calculate the time scale $\Delta t$ for the evaporation process
\eq{
\Delta t \propto M^{(1-b)/(1+b)} \left[= M^{d/(d-2)}\right]\,.
}{eq:ab007}
The second expression in \eqref{eq:ab007} refers to the Schwarzschild BH in $d+1$ space-time dimensions. The result \eqref{eq:ab007} provides a good estimate of the time-scale associated with BH evaporation, as long as the semi-classical approximation remains valid throughout most of the process. Typically, this requires that the BH initially have large $\mass$.

\subsubsection{Equation of State and Scaling Properties}

For some purposes it is useful to extract an equation of state $\bar{\psi}_c(T_c,\bar{D}_c)$ from the formulas above. It can be derived from \eqref{eq:ab20} upon expressing the right hand side as a function of $\bar{D}_c$ and $T_c$. To this end one has to find $M$ or $T$ as a function of $\bar{D}_c$ and $T_c$ using \eqref{eq:ab3}, \eqref{eq:ab4} and the defining relation \eqref{Tc}. As we shall demonstrate, for given $T_c$ and $\bar{D}_c$ there is in general an ambiguity, i.e., the BH mass $\mass$ or, equivalently, its Hawking temperature $T$ does not follow uniquely from specifying these quantities [cf.~the discussion around \eqref{ClassicalSolutions}]. A simple exception is provided by Rindler ground state models, where $T$ from \eqref{eq:ab3} not only is unique but actually is constant, thus establishing
\eq
{
{\rm Rindler:}\qquad \bar{\psi}_c(T_c,\bar{D}_c)= \frac{1}{a+1}\,4\pi T_c \,\bar{D}_c^{(1-a)/(1+a)} + \frac{a}{a+1}\,\frac{1}{4\pi T_c}\,\bar{D}_c^{(a-1)/(a+1)}-1\,.
}{eq:EOS1}
The dilaton chemical potential becomes minimal at $T_c=\sqrt{a}\bar{D}_c^{(a-1)/(a+1)}/(4\pi)$.
For the Witten BH $a=1$ the dilaton chemical potential depends on $T_c$ only and attains its minimum in the asymptotic region.  Also models with $b=1$ lead to a unique $T$ and to a simple equation of state,
\eq{
 \bar{\psi}_c(T_c,\bar{D}_c)=\frac{2}{a+2}\left(1+4\pi^2T_c^2\bar{D}_c^{-2a/(a+2)}\right)^{1/2}+\frac{a}{a+2} \left(1+4\pi^2T_c^2\bar{D}_c^{-2a/(a+2)}\right)^{-1/2}-1\,.
}{eq:EOS2}
For the JT model $a=0$ the dilaton chemical potential becomes independent from the dilaton charge. The same happens for all (A)dS ground state models
\eq{
(A)dS:\qquad \bar{\psi}_c(T_c)=\frac{1+b}{2}\,\frac{T_c}{k(T_c)}+\frac{1-b}{2}\,\frac{k(T_c)}{T_c}-1\,.
}{eq:EOS7}
However, the function $k(T_c)$ in general has more than one branch and is known implicitly only. For all other models one has to solve
\eq{
T=T_c \,\bar{D}_c^{(1+b-a)/(1+b+a)}\sqrt{1-\bar{D}_c^{-2(1+b)/(1+b+a)}\left(\frac{4\pi T}{1+b}\right)^{1+1/b}}
}{eq:EOS3}
for $T$ in terms of $T_c$ and $\bar{D}_c$. The solution of \eqref{eq:EOS3} is not unique in general. For instance, the Schwarzschild BH in 5 space-time dimensions, $a=-2b=2/3$, leads to two branches $T_\pm=T_c\big(1\pm\sqrt{1-144\pi^2/(T_c^4\bar{D}_c)}\,\big)^{1/2}/\sqrt{2}$. If we restrict the discussion to an equation of state for the asymptotic region $\bar{D}_c\to\infty$ it suffices to solve \eqref{eq:EOS3} perturbatively to leading order, in which case there is only one branch. For Minkowski ground state models the result is
\eq{
{\rm Minkowski \,\,ground \,\,state:}\qquad\lim_{\bar{D}_c\to\infty}\bar{\psi}_c (T_c,\bar{D}_c) \propto \frac{T_c^{2+2/b}}{\bar{D}_c^2}
}{eq:EOS4}
while for all other models it reads
\eq{
{\rm Others:}\qquad \lim_{\bar{D}_c\to\infty}\bar{\psi}_c (T_c,\bar{D}_c) \propto T_c^{1+1/b} \, \bar{D}_c^{-(1+b)(a+b-1)/(b(a+b+1))}\,.
}{eq:EOS5}
The proportionality constants in these equations are numerical coefficients depending on $a,b$. Asymptotically the equation of state for an ideal gas, $\bar{\psi}_c\propto T_c/\bar{D}_c$, arises in the limit $b\to\infty$, only. As a cautionary remark we recall that the asymptotic limit is not accessible for $b<0$ because the asymptotic specific heat \eqref{eq:MGS1784} is negative. By contrast, the near horizon approximation is always accessible. It corresponds to the limit $T_c\to\infty$ and allows to derive a simple equation of state for the branch where $T$ from \eqref{eq:EOS3} is finite at the horizon,
\eq{
{\rm Near \,\,horizon:}\qquad \bar{\psi}_c(T_c,\bar{D}_c) \approx \frac{4\pi}{a+b+1}\,T_c\,\bar{D}_c^{(1-a-b)/(a+b+1)}\,.
}{EOS6}

\TABLE{
\begin{tabular}{|>{$}l<{$}||c|c|c|c|c|c|c|}
\hline
$Quantity$ & Witten & JT & (A)dS & MGS & Rindler & $a=0$ & generic \\ \hline
\bar{D}_c,E_c,F_c,\bar{G}_c,\bar{\enthalpy}_c & ext. & ext. & ext. & ext. & ext. & ext. & ext. \\
\bar{\psi}_c,\bar{\isoth} & int. & int. & int. & int. & int. & int. & int. \\
S,C_D,\bar{C}_\psi       & ext. & ext. & ext. & ---  & ---  & ---  & ---  \\

T_c                      & int. & int. & int. & ---  & ---  & ---  & ---  \\
\sqrt{\killing_c}        & int. & ext. & ---  & int. & ---  & ext. & ---  \\
T                        & int. & ext. & ---  & ---  & int. & ---  & ---  \\
M                        & ext. & ---  & ---  & ext. & ---  & ---  & ---  \\
\hline
\end{tabular}
\caption{Extensitivity properties based upon the dilaton charge \eqref{eq:dc1}}
\label{tab:1}
}
Extensitivity properties of BHs are quite unusual; generically, the temperature is not intensive and depending on the choice of the dilaton charge, which is extensive by definition, either entropy is non-extensive or internal energy or both. For \eqref{DilaCharge} the dilaton field $D_c=X_c$ is extensive by definition and thus also entropy is extensive. However, generically no other quantity is extensive or intensive. For \eqref{eq:dc1} all thermodynamical potentials are extensive and the dilaton chemical potential becomes intensive. However, generically entropy is not extensive. We summarize the extensitivity properties with respect to \eqref{eq:dc1} of all $ab$-models in table \ref{tab:1}. Extensive (intensive) quantities are denoted by ``ext.'' (``int.''), while quantities which are neither are denoted by ``---''.

\TABLE{
\begin{tabular}{|>{$}l<{$}||c|c|c|c|c|c||>{$}c<{$}|>{$}c<{$}|c|}
\hline
           & Conf. & Scale & Shift & Ext. & Tol. & $G_2^{-1}$ & $Sign$ & $Sign$_{GS} & $D_c$-dep. \\ \hline
e^Q        & {\bf 2} & {\bf 1} & 0 & ---  & 0    & 0          & +      & +           & no  \\
w          & 0     & {\bf 1} & {\bf -1} & --- & 0 & 0         & +      & +           & no  \\
\bar{D}_c  & -1    & 0     & ---   & {\bf 1} & 0 & 1          & +      & +           & {\bf yes} \\
\killing_c & 2     & 2     & 0     & ---  & \ul{{\bf -2}} & 0 & +      & +           & no  \\
\Gamma_c   & 0     & 0     & ---   & ---  & 0    & {\bf 1}    & ?      & 0           & no  \\
M          & 0     & 1     & 1     & ---  & \ul{0} & 0        & \boldsymbol{+}      & \boldsymbol{0}           & no  \\\hline
T          & 0     & 1     & 0     & ---  & \ul{0} & 0        & +      & 0           & no  \\
T_c        & -1    & 0     & 0     & ---  & \ul{1} & 0        & +      & 0           & no  \\
F_c        & -1    & 0     & ---   & 1    & 1    & 1          & ?      & 0           & no  \\
S          & 0     & 0     & 0     & ---  & \ul{0} & 1        & +      & 0           & no  \\
E_c        & -1    & 0     & ---   & 1    & 0    & 1          & +      & 0           & no  \\
C_D        & 0     & 0     & 0     & ---  & -2   & 1          & ?      & 0           & no  \\
T_c S      & -1    & 0     & 0     & 1    & \ul{1} & 1        & +      & 0           & no  \\
M / w      & 0     & 0     & ---   & 0    & 0    & 0          & +      & 0           & no  \\ \hline
\bar{\psi}_c & ?   & 0     & ---   & 0    & 1    & 0          & ?    & 0           & yes \\
\bar{\isoth} & ?   & 0     & ---   & 0    & -1   & 0          & ?    & ?           & yes \\
\bar{G}_c  & ?     & 0     & ---   & 1    & \ul{1} & 1        & ?      & 0           & yes \\
\bar{\enthalpy}_c & ?  & 0 & ---   & 1    & \ul{1} & 1        & ?    & 0           & yes \\
\bar{C}_\psi & 0   & 0     & 0     & ---  & \ul{0} & 1        & ?    & 0           & yes \\
\bar{C}_\psi-C_D & 0 & 0   & 0     & ---  & 0    & 1          & ?    & 0           & yes \\
\hline
\end{tabular}\\
\caption{Summary of scaling properties}
\label{tab:2}
}
Besides extensitivity properties there are additional interesting scaling properties of thermodynamical quantities, which are summarized for generic models in table \ref{tab:2}. Here is an explanation of all abbreviations: ``Conf.'' refers to the conformal weight ($0$ means the quantity is conformally invariant and ``?'' means it transforms inhomogeneously); ``Scale'' refers to the scale ambiguity $e^Q\to ce^Q$, $w\to cw$ inherent to the definitions of these functions ($0$ means again that the quantity is independent from such rescalings); similarly, ``Shift'' refers to the shift ambiguity $w\to w+w_0$ (quantities which have no well-defined shift-weight are denoted by ``---''); ``Ext.'' denotes extensitivity properties ($1$ means extensive and $0$ means intensive); ``Tol.'' refers to the scaling with powers of the Tolman factor $1/\sqrt{\killing_c}$ close to the horizon (if it is positive the corresponding quantity diverges at the horizon and if the entry is underlined the whole cut-off dependence comes from a single Tolman factor with corresponding weight); ``$G_2^{-1}$'' determines the scaling with the gravitational coupling constant in front of the action; ``Sign'' is $+$ if the quantity is positive everywhere outside the horizon and ``?'' if the quantity may change its sign; ``Sign$_{GS}$'' is the same as ``Sign'' but for the ground state solution (therefore many quantities vanish); the last entry just keeps track of quantities which depend in an essential way on the definition of the dilaton charge. All scalings act multiplicatively [$s(A)\cdot s(B)=s(AB)$] except for ``Shift'' which acts additively [$s(A)+s(B)=s(A+B)$]. The bold entries in this table are definitions rather than results; e.g.~the quantity $e^Q$ is the conformal factor in front of the metric and thus by definition has a conformal weight of $+2$.

Finally, we collect various restrictions on the parameters from thermodynamical considerations. We recall that we have assumed $B>0$ for positive $w$, $M\geq 0$ for non-negative mass and $b>-1$ for a well-defined (positive) temperature. If we want space-time to be regular in the asymptotic region we encounter the inequalities  $a\leq 2$ and $a+b\leq 1$. If we demand that the dilaton chemical potential be non-negative everywhere then we need $0\leq a\leq 1+b$. Finally, if we wish the specific heat to be asymptotically positive the inequality $b\geq 0$ must hold.
These results are summarized in figure \ref{fig:ab}. The red (dark gray) region is excluded because $b<-1$ there, thus violating \eqref{wAsymptotic}. The orange (medium gray) region exhibits a curvature singularity in the ``asymptotic'' region $X_c\to\infty$. In the green (light gray) region the dilaton chemical potential can become negative. In the region below the $a$-axis the specific heat is asymptotically negative. The white region above the $a$-axis is free from thermodynamical pathologies regardless of the location of the cavity. The points S (W) [JT] correspond to the Schwarzschild BH (Witten BH) [JT model]. The dashed line denotes (A)dS ground state models and the dotted line Minkowskian ground state models. The $a$-axis contains all Rindler ground state models.
\FIGURE{
\includegraphics{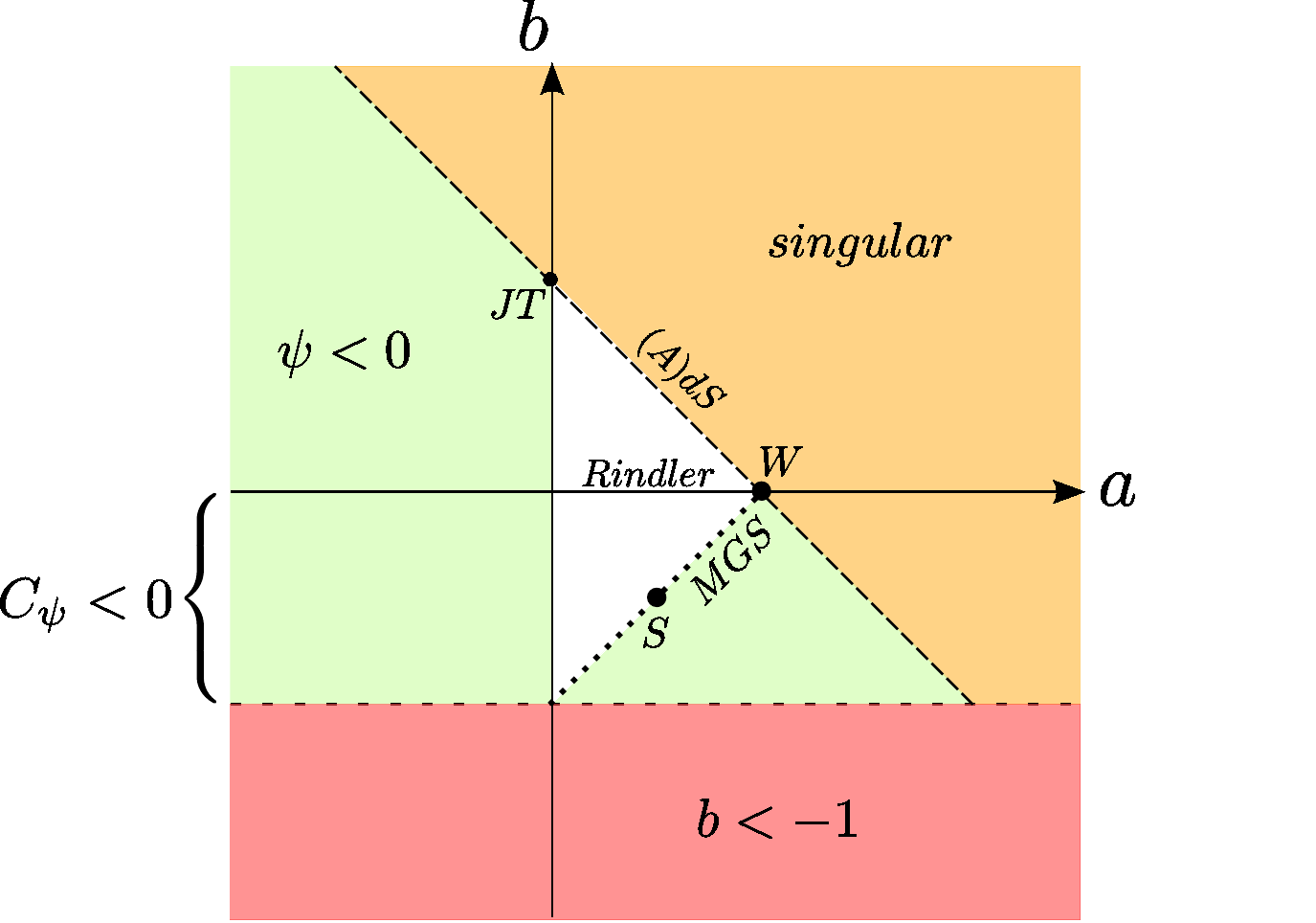}
\caption{Summary of the $ab$-family}
\label{fig:ab}
}

Two particular models of this class deserve further study. The Schwarzschild BH will be addressed in more detail in section \ref{sec:HigherDimensionalExamples} and the Witten BH will be studied in the next section.

\section{Examples from String Theory}
\label{sec:StringExamples}

In this section we consider two-dimensional BHs that emerge as solutions, either approximate or exact, of string theory. In \cite{Witten:1991yr} it was shown that the Euclidean BHs studied in \cite{Elitzur:1991cb} 
admit an exact CFT description in terms of an $SL(2,\BR)/U(1)$ gauged WZW model. In the large $k$ limit, where $k$ is the level of the $SL(2)$ current algebra, the corresponding background takes the form
\begin{equation}\label{eq:WittenBH}
   ds^2 = \alpha' \, k \left(dy^2 + \tanh^2{y} \,d\theta^2 \right)  \qquad X = X_0\,\cosh^{2}{y} ~.
\end{equation}
This solution is commonly referred to as the `Witten BH'. It is related\footnote{The precise relationship is as follows: in the limit $k\to\infty$ (or, equivalently, $p\to 0$) upon identifying $y=bx$ and $\tau=\sqrt{\alpha^\prime k}\,\theta$ the line-element and dilaton \eqref{eq:WittenBH} emerge from \eqref{ExactStringBH}.} to the exact background obtained in \cite{Dijkgraaf:1992ba}
\begin{equation}\label{ExactStringBH}
   ds^2 = dx^2 + \frac{\tanh^{2}(b\,x)}{1-p\,\tanh^{2}(b\,x)}\,d\tau^2 \qquad
   X= X_0 \, \cosh^{2}(b\,x)\,\sqrt{1-p\,\tanh^{2}(b\,x)}
\end{equation}
where the parameters $b$ and $p$ are
\begin{equation}\label{eq:EBH3}
   \alpha'\,b^{2} = \frac{1}{k-2} \qquad p \defeq \frac{2\,\alpha'\,b^{2}}{1+2\,\alpha'\,b^{2}}
    = \frac{2}{k} ~.
\end{equation}
The background \eqref{ExactStringBH} can be expressed in the form \eqref{metric} by means of coordinate transformations described in \cite{Grumiller:2005sq}.

In the following subsections we wish to consider BHs for the full range $k \in (2,\infty)$ allowed by the CFT. However, in order for the background \eqref{ExactStringBH} to be a solution of string theory it must satisfy the condition
\begin{equation}\label{CentralCharge}
    \textrm{Dim} - 26 + 6\,\alpha'\,b^2 = 0 ~.
\end{equation}
Because the target space here is two-dimensional, $\textrm{Dim}=2$, requiring the correct central charge fixes the level at the critical value $k_{\rm crit} = 9/4$. Following \cite{Kazakov:2001pj}, we vary $k$ by allowing for additional matter fields that contribute to the total central charge, modifying the condition \eqref{CentralCharge} so that $k \neq 9/4$.

\subsection{Witten Black Hole}
\label{sec:WittenBH}

The Witten BH has the most intriguing thermodynamics of all the models we have examined so far \cite{Gibbons:1992rh, Nappi:1992as, Davis:2004xi}. We focus on the background \eqref{eq:WittenBH} as a solution of the string theory $\beta$-functions at lowest order in $\alpha'$. The $\beta$-functions can be derived from a target-space action of the form \eqref{Action} with
\begin{equation}
    U = - \frac{1}{X} \quad \quad V = - \frac{\lambda^2}{2} \, X
\end{equation}
where $\lambda = 4/\sqrt{\alpha^\prime}$. With these potentials and convenient choices for the integrations constants in \eqref{QDef} and \eqref{wDef} the functions $Q$ and $w$ are given by
\eq{
e^Qw=1\,,\qquad w = \la X\,.
}{eq:WBH}
Because $w$ is linear in $X$, the surface gravity and asymptotic temperature $T=\la/(4\pi)$ do not depend on the mass $\mass$. As a result, the condition \eqref{ClassicalSolutions} identifies a single solution consistent with the boundary conditions $\beta_c$ and $D_c$. The relation $2M=\la X_h$ then implies a simple proportionality between mass and entropy
\begin{equation}
    M = T \, S ~.
\end{equation}
An immediate consequence is that the Witten BH satisfies the Gibbs-Duhem relationship: $G^W_c=\bar{G}^W_c=0$.

The two definitions for the dilaton charge \eqref{DilaCharge} and \eqref{eq:dc1} are both linear in $X_c$ for the Witten BH. They are related by $\bar{D}_c = \lambda \, D_c$. Therefore, the dilaton chemical potential associated with $\bar{D}_c$,
\eq{
\bar{\psi}_c^W=\frac12 \left(1-\frac{X_h}{X_c}\right)^{1/2}+\frac12 \left(1-\frac{X_h}{X_c}\right)^{-1/2}-1\,,
}{eq:WBH1}
is related to $\psi_c^W$ from \eqref{eq:td7} by the constant rescaling $\psi_c^W = \lambda \, \bar{\psi}_c^W$. Both the internal energy and the Helmholtz free energy
\eq{
E_c^W=\la\, X_c\left(1-\sqrt{1-\frac{X_h}{X_c}}\right) \quad \quad F_c^W=E_c^W-T_c^WS
}{eq:WBH2}
have finite asymptotic limits, $E^W\to M$ and  $F^W\to 0$, respectively. The extensitivity properties of the Witten BH are comparable to those of the standard thermodynamics of ideal gases, as can be seen from table \ref{tab:1} on page \pageref{tab:1}.

There are various reasons to expect that a semi-classical analysis of the Witten BH might encounter problems. For instance, the curvature is of order $1/\alpha'$ near the horizon, indicating that corrections have to be taken into account. It is not surprising that this system also exhibits some peculiar thermodynamic properties. The most obvious is the behavior of the the specific heat, which is given by
\eq{
C_D^W=4\pi(X_c-X_h) ~.
}{eq:WBH3}
The specific heat is positive for all $X_c > X_h$, but it diverges linearly if we try to remove the cavity wall to infinity. In addition, the quantities $C_\psi$ and $\isoth$ are not well-defined. However, since we know that the Witten BH is related to the exact solution \eqref{ExactStringBH}, it is reasonable to expect that these problems might be absent in a more complete treatment. As we shall demonstrate in the next subsection, $\alpha^\prime$ corrections have a significant impact on the thermodynamics.

\subsection{String Theory Is Its Own Reservoir: Exact String Black Hole}
\label{sec:ESBH}

The discussion of the canonical ensemble in section \ref{sec:Thermo} required the addition of a cavity wall at $X_c$ that couples the system to a thermal reservoir. In most cases thermodynamic stability is only possible for finite values of $X_c$. We have not questioned the nature of this reservoir, or the form of the coupling between the two systems. In previous examples we simply assumed that it was possible to couple a given model to some other degrees of freedom that functioned as the reservoir. String theory is fundamentally different in this regard. It is not possible to couple arbitrary degrees of freedom to the theory, or place an abrupt cut-off on the space-time fields.
Therefore, if we apply our methods to a non-compact solution of string theory any cut-offs introduced during a calculation \emph{must} be removed.

As shown in the previous subsection, the specific heat of the Witten BH diverges as the cut-off is removed to infinity. Remarkably, this and other issues are resolved when $\alpha^\prime$ corrections are taken into account. The metric and dilaton for the ESBH were given in \eqref{ExactStringBH}. A target space action of the form \eqref{Action} was constructed for the Exact String BH in \cite{Grumiller:2005sq}. In our current notation the relevant functions determining the model are given by\footnote{Winding/momentum mode duality relates the Exact String BHs to a different set of target space geometries, namely the Exact String Naked Singularities. They are also described by \eqref{eq:EBH1} and \eqref{eq:EBH2}, but with sign changes in front of the square root and in front of the $\arcsinh$.}
\eq{
e^Qw=1\,,\qquad w(X) = 2 b \,\left(\sqrt{\rho^2+1}+1\right)\,,
}{eq:EBH1}
where $b$ was defined in \eqref{eq:EBH3}, and the canonical dilaton $X$ is related to a new field $\rho$ by
\eq{
X=\rho+\arcsinh{\rho} ~.
}{eq:EBH2}
It will generally be more convenient to work with the field $\rho$, rather than $X$. The relation \eqref{eq:EBH2} between the two fields is globally invertible, and the limits $X\to\infty$ and $X\to 0$ coincide with the corresponding limits for $\rho$. Therefore, $\rho$ can be employed not only locally but globally. Expressed in terms of $\rho$, the square of the Killing norm is given by
\eq{
\killing = 1-\frac{k}{\sqrt{\rho^2+1}+1} ~.
}{eq:EBH8}
This vanishes at
\eq{
\rho_h=\sqrt{k(k-2)}\,.
}{eq:EBH9}
which establishes the location of the horizon in terms of the level $k$.

As in \cite{Grumiller:2005sq}, the integration constant in $w(X)$ has been chosen assuming a Minkowski ground state, i.e. $e^{Q} w = 1$.
Comparing the general form of the Killing norm \eqref{xiDef} with \eqref{eq:EBH8} shows that the mass parameter is then proportional to the level: $\mass^{MGS} = bk$.
This means that the choice of $w_0$ in \eqref{eq:EBH1} identifies the ground state with $k=0$. However, the CFT description requires $k>2$. This can also be seen from the Hawking temperature \cite{Kazakov:2001pj},
\eq{
T^{ESBH} = \frac{b}{2\pi}\sqrt{1-\frac{2}{k}} \,,
}{eq:EBH5}
which is real and positive only if $k > 2$. Therefore, we must shift the constant $w_0$ so that the ground state corresponds to level $k=2$. This is accomplished by replacing $w$ with $w-4b$ in \eqref{eq:EBH1}. Notice that this is the largest shift by a negative constant that is consistent with $w > 0$. The result is
\eq{
e^{Q^{ESBH}(X)} = \frac{1}{2b}\,\frac{1}{\sqrt{\rho^2+1}+1}\,,\qquad w^{ESBH}(X)=2b\left(\sqrt{\rho^2+1}-1\right) ~.
}{eq:EBH6}
Comparing the resulting expression for the Killing norm with \eqref{eq:EBH8} gives the mass parameter in terms of $b$ and $k$
\eq{
\mass^{ESBH}=b(k-2)\,.
}{eq:EBH4}
With \eqref{eq:EBH6} the dilaton charge \eqref{eq:dc1} reads
\eq{
\bar{D}_c^{ESBH}=2 b\,\rho_c\,.
}{eq:EBH12}
For given $\beta_c$ and $D_c$ the equation \eqref{ClassicalSolutions} has a unique solution with level $k>2$. In the asymptotic region $D_c\to\infty$ the temperature is bounded from above and below for $k\in(2,\infty)$ according to \eqref{eq:EBH5}. As we shall show in the next subsection, the thermal partition function diverges in the $k \to \infty$ limit, which identifies the maximal temperature $T_H=b/(2\pi)$ as the Hagedorn temperature. As might have been anticipated on general grounds the asymptotic temperature has to be positive and below the Hagedorn temperature for the existence of the ESBH solution.

\subsubsection{Thermodynamics of the Exact String Black Hole}

With \eqref{eq:EBH6} as a starting point we can calculate all thermodynamical quantities of interest for the ESBH by applying the general results of section \ref{sec:Thermo}. The cut-off is always removed at the end of a calculation, so we drop the subscript `c' from the following expressions. The result for the entropy,
\eq{
S^{ESBH}=2\pi \left(\sqrt{k(k-2)} + \arcsinh{\sqrt{k(k-2)}}\right)\,,
}{eq:ESBH10}
agrees with equation (4.17) found in \cite{Grumiller:2005sq} (In that reference $M=k/2$.). For $k=9/4$ we obtain $S^{ESBH}=2\pi(3/4+\ln{2})$. In the large $k$ limit the entropy $S^{ESBH}=2\pi k + {\mathcal O} (\ln{k})$ asymptotes to the result for the Witten BH with logarithmic $\alpha^\prime$ corrections \cite{Grumiller:2005vy}. The free energy,
\eq{
F^{ESBH}=-b\,\sqrt{1-\frac{2}{k}}\,\arcsinh{\sqrt{k(k-2)}}
}{eq:EBH10}
is manifestly non-positive~\footnote{It is worth mentioning that there is also a CDV solution with $X=0$ and AdS geometry with curvature proportional to $b^2$. It has vanishing free energy. Therefore the Exact String BH is stable against tunneling (cf.~subsection \ref{sec:CDV}).}. This would not have been the case with the original choice of $w_0$ that lead to \eqref{eq:EBH1}. The result \eqref{eq:EBH10} coincides with (5.7) in \cite{Grumiller:2005sq}. The internal energy is given by the mass, $E^{ESBH}=\mass^{ESBH}=b(k-2)$. The specific heat
\eq{
C_D = \bar{C}_\psi = 2\,\pi\, k \,\sqrt{k(k-2)}
}{eq:EBH11}
is positive and finite. Thus the problem encountered for the Witten BH is absent: removing the cut-off is possible for the Exact String BH and leads to a positive and finite specific heat. It is worth mentioning that the specific heat and entropy both vanish in the limit $T\to 0$, so the Exact String BH is one of the few examples for which the third law of thermodynamics holds.

Even though the Minkowski ground state condition is not satisfied by \eqref{eq:EBH6}, the Killing norm still approaches unity as the cut-off is removed. Therefore the asymptotic limit of \eqref{eq:dc3} is constant, $\eta^{ESBH}=1/2$. As a consequence the dilaton-chemical potential vanishes as $X_{c}^{-2}$ in the limit $X_c \to \infty$. The Gibbs-Duhem relation
\eq{
\bar{G}^{ESBH} = F^{ESBH} = \mass^{ESBH}-T^{ESBH}S^{ESBH} \neq 0
}{eq:EBH13}
is violated even in the large $k$ limit,
\eq{
\lim_{k\to\infty} \bar{G}^{ESBH} = \lim_{k\to\infty} F^{ESBH} = -b\ln{2k} + \mathcal{O}\left(\frac{1}{k}\,\ln{k}\right)\,.
}{eq:EBH18}
Thus, comparing the thermodynamic behavior of the exact String BH with that of the Witten BH demonstrates the importance of $\alpha'$ corrections, even in the large $k$ limit. In the limit $k\to 2$ the Gibbs-Duhem relation,
\eq{
\lim_{k\to 2}\bar{G}^{ESBH} = \lim_{k\to 2} F^{ESBH} = -b(k-2) + \mathcal{O}(k-2)^2\,,
}{eq:EBH19}
is restored.

It is interesting to observe how the free energy scales with the level $k$ in various limits. For large values of $k$ the free energy scales logarithmically, while for $k$ close to the ground state value it is linear in $k-2$. This implies corresponding behavior of the thermal partition function \eqref{PartitionFunction2}. Suppose that we fix the factor in front of the action as $\couplingESBH/(2\pi)$, where $\couplingESBH$ is some positive parameter, so that the curvature term in the action \ref{Action} is given by $-\couplingESBH XR/(4\pi)$. Then the thermal partition function in the large $k$ limit is given by
\eq{
\left. \ZZ^{ESBH}\raisebox{14pt}{}\right|_{k \gg 2} = (2k)^\couplingESBH + \dots \sim \left(S^{ESBH}\right)^\couplingESBH + \dots
}{eq:EBH20}
In this limit the partition function grows as a positive power of $k$. It diverges in the $k \to \infty$ limit, as the temperature approaches the asymptotic value $T_H=b/(2\pi)$. In this sense, $T_H$ represents the Hagedorn temperature. Similarly, for $k\to 2$ we obtain
\eq{
\left. \ZZ^{ESBH}\raisebox{14pt}{}\right|_{k \to 2} = e^{\couplingESBH\sqrt{2k-4}}+\dots \sim e^{\frac{\couplingESBH}{4\pi}\,S^{ESBH}}+\dots
}{eq:EBH21}
Thus, in the weak coupling limit the partition function grows only with a power of entropy, while in the strong coupling limit it grows exponentially with entropy. This is consistent with fermionization in the strong coupling regime.\footnote{\label{fn:1} In the WZW formulation the stress tensor is given by
\[
T(z)=\frac{1}{k-2}\eta_{ab}J^aJ^b + \frac{k}{4}(\partial\phi)^2 +b\partial c\,.
\]
Here $\eta_{ab}$ is the metric on the algebra $sl(2,\mathbb{R})$ and the $J^a$ are given by equation (2.11) in \cite{Dijkgraaf:1992ba}. For $k\to 2$ the first term dominates and the system behaves like a free Fermi gas, cf.~e.g.~\cite{Ford:2000nc}. 
This suggests that an appropriate interpretation of the Exact String BH in that limit is not in terms of some effective geometry but rather in terms of free fermions.}
For arbitrary values of $k$ we obtain
\eq{
\ZZ^{ESBH}(k) = e^{\couplingESBH\,\arcsinh{\sqrt{k(k-2)}}}\,.
}{eq:EBH22}
Because $\exp{[\arcsinh{(3/4)}]}=2$ we obtain the interesting result
\eq{
\ZZ^{ESBH}(9/4) = 2^\couplingESBH\,.
}{eq:EBH23}
As a caveat, we remind the reader that these results apply to the semi-classical approximation of the thermal partition function, based on the action for the ESBH given in \cite{Grumiller:2005sq}. They do not take into account contributions from the tachyon that appears in the bosonic string spectrum, or from any matter that must be coupled to the ESBH to maintain the correct central charge \eqref{CentralCharge}. Furthermore, while we have focused on the effects of $\alpha'$ corrections, we have not taken into account contributions from world-sheet topologies other than the leading term.

\subsubsection{Thermodynamical Derivation of the Exact String Black Hole}

It is interesting to note that the potentials \eqref{eq:EBH1} can be derived purely from thermodynamical considerations. Of course, this requires some ad-hoc assumption: we look for a model which may be shifted to a Minkowski ground state model and whose BH solutions have a specific heat whose asymptotic behavior is
\eq{
C_D=\bar{C}_\psi\propto \mass^2 T
}{eq:EBH14}
for all values of $\mass$ and $T$. The rationale behind this assumption comes from the consideration of the weak ($k\to\infty$) and strong ($k\to 2$) coupling limit. In the former case we have seen that the ESBH is approximated by the Witten BH which has a divergent asymptotic specific heat. Taking into account corrections from fluctuations at 1-loop level leads to a specific heat $C_D\propto M^2T$ \cite{Grumiller:2003mc}. This is a non-stringy calculation, but as long as we are not interested in the precise proportionality factor it should not matter which kind of fluctuations we consider perturbatively: thermal fluctuations, $\alpha^\prime$ corrections and quantum fluctuations of some additional massless matter fields are expected to yield the same qualitative behavior. In the strong coupling limit the system can be fermionized (cf.~footnote \ref{fn:1}). A free Fermi gas has a specific heat of the form $C\propto T$. The proportionality constant can be rescaled such that it includes a factor $\mass^2$. Then the weak and strong coupling limits qualitatively both yield the same result for the specific heat, which motivates our Ansatz \eqref{eq:EBH14}.

With $C_D\propto w_h'/w''_h$, $T\propto w_H'$ and $2M=w_h$ the most general model \eqref{Action} obeying \eqref{eq:EBH14} has to fulfill the autonomous differential equation
\eq{
w''\propto\frac{1}{w^2}\,.
}{eq:EBH15}
Without loss of generality we fix the proportionality constant in \eqref{eq:EBH15} to unity. Its general solution is
\eq{
\int\frac{d w}{\sqrt{c_1-2/w}}=c_0\pm X\,.
}{eq:EBH16}
The integration constant $c_1$ may be adjusted by rescaling $w$, $c_0$ and $X$, so we fix it to unity. The integration constant $c_0$ may be adjusted by shifting the origin of $X$, so we fix it to zero. The sign ambiguity is resolved by requiring $X$ to be positive. Thus we obtain
\eq{
\sqrt{w(w-2)}+\ln{\left[w-1+\sqrt{w(w-2)}\right]} = X\,.
}{eq:derESBH3}
Defining $\rho=\sqrt{w(w-2)}$ reveals that \eqref{eq:derESBH3} actually is equivalent to \eqref{eq:EBH2}. Solving $w$ in terms of $\rho$ and taking the branch of positive $w$ leads to
\eq{
w(X)=\sqrt{\rho^2+1}+1\,,\qquad X=\rho+\arcsinh{\rho}\,.
}{eq:EBH17}
This is the same as $w(X)$ in \eqref{eq:EBH1} for $b=1/2$. If we had chosen a different proportionality constant in \eqref{eq:EBH15} we would obtain a different value for $b$ but otherwise the same result for $w$. The function $Q$ is fixed by the Minkowski ground state property \eqref{eq:MGS1}. Thus, the Exact String BH in the Minkowski ground state representation \eqref{eq:EBH1} is the most general two-dimensional dilaton gravity model with a Minkowskian ground state that has an asymptotic specific heat $C_D=\bar{C}_\psi\propto M^2 T$.

\section{Spherically Symmetric Black Holes in Higher Dimensions}
\label{sec:HigherDimensionalExamples}

  Up to this point we have focused on theories in two dimensions, but the results of section \ref{sec:Thermo} can also be applied to solutions of gravitational theories in higher dimensions, as long as the on-shell action reduces to the form \eqref{CutOffAction} or \eqref{eq:max5}. With a few caveats, this leads to a simple description of the thermodynamics of a class of spherically symmetric BHs in $d+1 > 2$ dimensions, with or without a cosmological constant.

Pure gravity with a cosmological constant in $d+1 > 2$ dimensions is described by the action
\begin{equation}\label{DAction}
   I_{d+1} = - \frac{1}{16\pi G_{d+1}} \int_{\MM} \nts d^{\,d+1}x\,\sqrt{g_{d+1}}\,\left(R_{d+1}-2\,\Lambda\right)
    - \frac{1}{8 \pi G_{d+1}} \, \int_{\dM} \bns d^{\,d}x \, \sqrt{\gamma_{d}} \, \left(K_{d} +
    \ldots \right) ~.
\end{equation}
Subscripts `$d+1$' and `$d$' have been used to distinguish quantities from their two- and one-dimensional analogs. The `$\ldots$' in the boundary integral indicates boundary counterterms, whose precise definition depends on the space-times asymptotics \cite{Mann:2005yr,Papadimitriou:2005ii}.
We are interested in solutions of this theory that can be expressed as a direct product of the two-dimensional metric \eqref{metric} and the metric on a round $d-1$ sphere
\begin{equation}\label{DMetric}
   ds^2 = \xi(r) \, d\tau^2 + \frac{1}{\xi(r)}\,dr^2 + (G_{d+1})^{\frac{2}{d-1}}\,\varphi(r)^2 \, d\Omega_{d-1}^{\, 2} ~.
\end{equation}
Explicit factors of the Newton's constant have been included in this expression so that the field $\varphi(r)$ is dimensionless. Ignoring for a moment the contributions from the boundary counterterms, the on-shell reduction of \eqref{DAction} on the $d-1$ sphere leads to a two-dimensional expression of the form \eqref{Action}. The dilaton $X$ in this action is related to the scalar field $\varphi$ by
\begin{equation}\label{ReducedDilaton}
   X(r) = \Upsilon \,G_{d+1} \, \varphi(r)^{d-1} \qquad \Upsilon \defeq \frac{A_{d-1}}{8\,\pi\,G_{d+1}}
\end{equation}
where $A_{d-1}$ is the solid angle subtended by the $d-1$ sphere, and the constant $\Upsilon$ has been defined for convenience. Thus, the dilaton $X(r)$ is the proper area of a sphere with coordinate radius $r$, in $d+1$ dimensional Planck units. The kinetic and potential functions for the dilaton are
\begin{gather} \label{ReducedU}
 U(X) = - \left(\frac{d-2}{d-1}\right) \, \frac{1}{X}   \\ \label{ReducedV}
 V(X) = - \frac{1}{2} \, (d-1)(d-2)\, \Upsilon^{\frac{2}{d-1}}
    \, X^{\frac{d-3}{d-1}} + e\,\frac{d(d-1)}{2\,\ell^2} \, X ~.
\end{gather}
The last term in $V$ comes from the cosmological constant, which we parameterize in terms of a length scale $\ell$ by
\begin{equation}
    \Lambda \defeq e \frac{d(d-1)}{2\,\ell^2} \qquad e = \pm 1, 0 ~.
\end{equation}
The coefficient of the reduced action explicitly sets the two-dimensional Newton's constant to $8\pi G_2 = 1$. From \eqref{ReducedU} and \eqref{ReducedV} we obtain
\begin{gather} \label{ReducedQ}
e^{Q(X)} = \frac{\Upsilon^{\frac{1}{1-d}}}{d-1}\,X^{\frac{2-d}{d-1}}\,,\\
\label{Reducedw}
w(X) = (d-1)\,\Upsilon^{\frac{1}{d-1}}\,X^{\frac{d-2}{d-1}} \left( 1 -
    \frac{e}{\,\ell^2} \, \Upsilon^{-\frac{2}{d-1}} \,
    X^{\frac{2}{d-1}}\right) \,,
\end{gather}
where we have fixed the free constants in the definitions \eqref{QDef} and \eqref{wDef} conveniently.

The results of section \ref{sec:Thermo} can be applied to these solutions of the $d+1$ dimensional theory. This is because the on-shell action, after integrating over the sphere, takes the form \eqref{CutOffAction}. Of course, the contributions from the boundary counterterms, alluded to by the `$\ldots$' in \eqref{DAction}, must be properly accounted for. Otherwise the reduced on-shell action would diverge, as in \eqref{DivergentAction}. This raises an interesting technical problem. The formulation of the boundary counterterms in the $d+1$ dimensional theory depends on the asymptotics of the space-time \cite{Kraus:1999di, Mann:2005yr}. This would seem to imply that the reduction of theories with $\Lambda = 0$ and $\Lambda \neq 0$ must be studied separately. We take a different approach. Rather than working out the reduction of the appropriate boundary counterterms in \eqref{DAction}, we assume that the two-dimensional boundary counterterm derived in subsection \ref{sec:BoundaryTerm} gives a sensible renormalization of the reduced on-shell action. For the solution described above, the boundary counterterm is
\begin{equation} \label{GeneralCT}
  I_{CT} = - \beta \, w(X) \, \sqrt{1 - \frac{2\,M}{w(X)}}\,.
\end{equation}
This counterterm gives the `correct' on-shell action in the sense that the analysis of section \ref{sec:Thermo} recovers the standard results associated with the thermodynamics of BHs whose metrics take the form \eqref{DMetric}. An immediate consequence, due to \eqref{ReducedDilaton}, is
\begin{equation}\label{}
  S = 2 \pi X_h = \frac{A_h}{4 G_{d+1}} ~.
\end{equation}
The entropy of these BHs is \emph{always} given by one-quarter of the proper area of the horizon, in $d+1$ dimensional Planck units.

There is no reason to believe that the counterterm \eqref{GeneralCT} can be lifted to higher dimension, to recover the counterterms of the original theory. The combination of the two-dimensional metric and scalar field are simply not sufficient to reproduce the appropriate functional dependence on the higher-dimensional metric. But despite this caveat, it seems remarkable that the quasi-local thermodynamics of spherically symmetric BHs with different space-time asymptotics can be recovered from the reduced on-shell action, renormalized by the two-dimensional counterterm \eqref{ICT}.

\subsection{Schwarzschild Black Hole in $d+1 \geq 4$ Dimensions}
\label{sec:ReducedSchwarzschild}

\begin{figure*}[htb!]
\centering%
\includegraphics{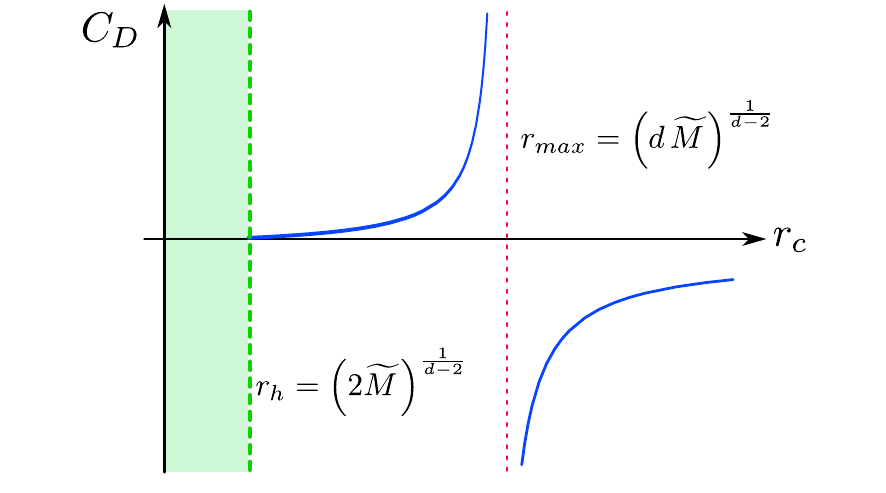}
\caption{Heat capacity for the Schwarzschild BH in dimension $d+1 \geq 4$.}
\label{fig:SchwarzschildHeatCapacity}
\end{figure*}

As a first example, consider the on-shell reduction of \eqref{DAction} with $\Lambda = 0$. Setting $\varphi \sim r$, up to factors of the Newton's constant, gives the standard form of the Schwarzschild solution
\begin{gather}
    \xi(r) = 1 - \frac{2  \widetilde{M}}{r^{d-2}} \\
    X(r) = \Upsilon \, r^{d-1}
\end{gather}
where the mass parameter $\widetilde{M}$ is related to the constant $M$ by
\begin{equation}
    \widetilde{M} \defeq \frac{M}{(d-1)\Upsilon} ~.
\end{equation}
In section \ref{sec:Thermo} the dilaton charge $D_c = X_c$ and proper periodicity $\beta_{c}$ were held fixed at $r=r_c$. This is precisely the canonical ensemble studied in \cite{Brown:1989fa, Brown:1991gb}, 
where the BH is enclosed inside a cavity of radius $r_c$ whose area and proper temperature are held fixed by means of an external thermal reservoir. The relation \eqref{ReducedDilaton} identifies the dilaton charge $D_c = X_c$ as the area of the cavity, so the dilaton chemical potential \eqref{eq:td7} is the corresponding pressure
\begin{equation}
    p_c = \frac{1}{2}(d-2)\,r_{c}^{-1}\,\left( \sqrt{\xi(r_c)} + \frac{1}{\sqrt{\xi(r_c)}} - 2 \right) ~.
\end{equation}
The internal energy of the system can be expressed as
\begin{equation}\label{SchwarzschildEnergy}
    E_c = (d-1) \, \Upsilon \,r_{c}^{\,d-2}\,
    \left( 1 - \sqrt{1 - \frac{2  \widetilde{M}}{r_{c}^{d-2}} } \right) ~.
\end{equation}
The asymptotic limit of this quantity gives the ADM mass $M$. Furthermore, the internal energy satisfies the quasi-local form of the first law \eqref{eq:td12},
\begin{equation}\label{}
  d E_{c}(S,A_c) = T_c \, dS - p_c \, d A_c~,
\end{equation}
where $A_c$ is the area of the cavity in Planck units ($8\pi G_{d+1}=1$).
Finally, the expression \eqref{eq:td18} for the heat capacity recovers the well-known result for the thermodynamic stability of the Schwarzschild BH enclosed by an isothermal cavity in d+1 dimensions
\begin{equation}
    C_{D} = -4\pi \, M \, r_h \, \frac{r_{c}^{\,d-2}-2\widetilde{M}}{r_{c}^{\,d-2}-d \,\widetilde{M}} ~.
\end{equation}
This quantity is positive for $r_h < r_c < r_{max}$, as indicated in figure \ref{fig:SchwarzschildHeatCapacity}.

\subsection{Asymptotically AdS Space-Times}
\label{sec:AAdS}

Now we turn our attention to the case of non-zero cosmological constant. For definiteness, we take the cosmological constant to be negative ($e=-1$) and consider asymptotically AdS space-times\footnote{Asymptotically de Sitter space-times with $\Lambda > 0$ can also be addressed in this framework, but many of the working assumptions made in sections \ref{sec:Action} and \ref{sec:Thermo} must be changed. We shall discuss this briefly in section \ref{sec:Discussion}.}. Setting $\varphi \sim r$ leads to the Schwarzschild-AdS solution
\begin{gather}\label{}
  \xi(r) = 1 + \frac{r^2}{\ell^2} - \frac{2\,\widetilde{M}}{r^{d-2}} \quad \quad  X = \Upsilon \, r^{d-1} ~.
\end{gather}
As in the previous case, holding $D_c = X_c$ and $\beta_{c}$ fixed at $r_c$ corresponds to an AdS BH enclosed by an isothermal cavity. The thermodynamics of this system was studied in \cite{Brown:1994gs, Akbar:2004ke}.

Rather than reproduce the full catalog of results regarding the Schwarzschild-AdS solution, we focus on a few interesting properties which are qualitatively different than the previous example. First, the internal energy \eqref{eq:td10} is given by
\begin{equation}\label{}
  E_c = \frac{(d-1)}{\ell}\,\Upsilon\,r_{c}^{d-1}\,\left( \sqrt{1+ \frac{\ell^2}{r_{c}^2}} -
    \sqrt{1+ \frac{\ell^2}{r_{c}^2} - \frac{2 \widetilde{M} \ell^2}{r_{c}^d}} \right) ~.
\end{equation}
Unlike the result \eqref{SchwarzschildEnergy}, the asymptotic limit of the internal energy vanishes due to red-shifting
\begin{equation}\label{}
  \lim_{r_c \to \infty} E_c = 0 ~.
\end{equation}
However, as in section \ref{sec:Energy}, the conserved charge is \cite{Brown:1991gb} 
\begin{equation}\label{ConservedChargeEnergy}
  Q_{\partial_{\tau}} \defeq \lim_{r_c \to \infty} \sqrt{\xi(r_c)} \, E_c = M ~.
\end{equation}
This is the same result obtained in higher dimensions using any one of a number of methods \cite{Hollands:2005wt}. But, notably, it does not agree with the standard AdS/CFT result when the space-time dimension $d+1$ is odd. In that case, global AdS (the ground state) has $Q_{\partial_{\tau}} = M_{cas}$, where $M_{cas}$ is a Casimir energy associated with the dual field theory on $S^{1} \times S^{d-1}$ \cite{Balasubramanian:1999re, Emparan:1999pm}. Reproducing the Casimir term is possible in the present context: an appropriate shift in $w_0$ can be used to shift the value of \eqref{ConservedChargeEnergy} by an amount $M_{cas}$, for all solutions. But from the two-dimensional point of view the shift is arbitrary. The counterterm \eqref{GeneralCT} is completely ignorant of the physics attached to this issue in the higher-dimensional theory.

\begin{figure*}[htb!]
\centering%
\includegraphics{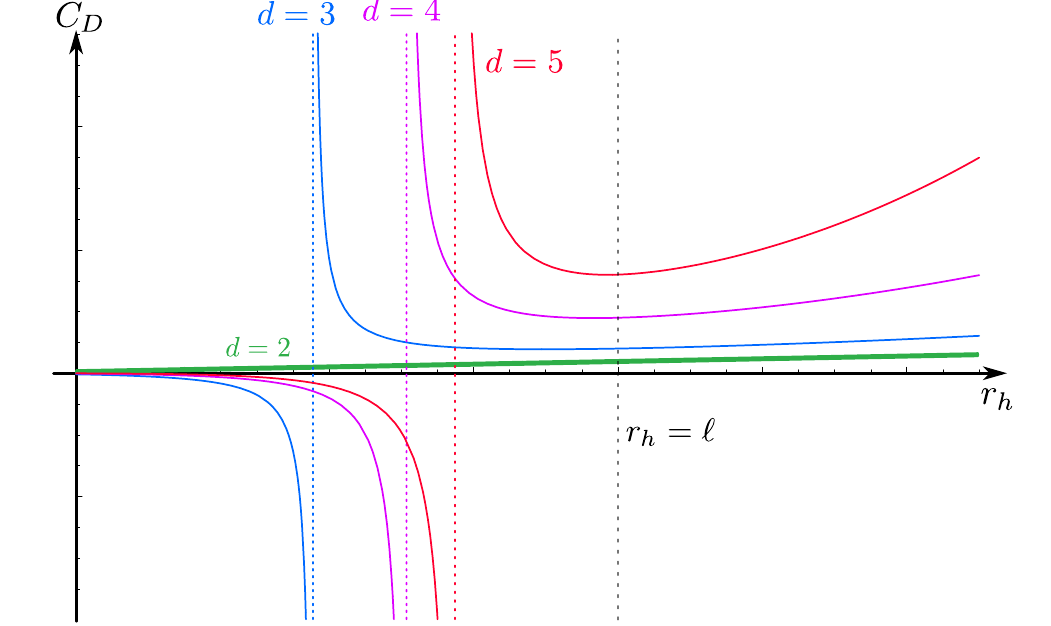}
\caption{Heat capacity for the Schwarzschild-AdS BH in $d+1$ dimensions, with the cavity wall sent to infinity. The horizontal axis is the horizon radius $r_h$, and the vertical axis is the $X_c \to \infty$ limit of $C_{D}$. }
\label{fig:AdSSchwarzschildHeatCapacity}
\end{figure*}

Another interesting result is the structure of the heat capacity for an asymptotically AdS BH. The result is, in general, very complicated for finite $r_c$ \cite{Akbar:2004ke}. But the presence of a cosmological constant can stabilize the BH even when the cavity wall is removed to infinity. Using \eqref{eq:td100}, the $r_c \to \infty$ limit of the heat capacity is
\begin{equation}\label{}
  \lim_{r_c \to \infty} C_{D} =  (d-1)\, 2 \pi X_h \, \left( \frac{d\,X_{h}^{\frac{2}{d-1}}+ (d-2)\,\ell^2\,\Upsilon^{\frac{2}{d-1}}}{d\,X_{h}^{\frac{2}{d-1}}-(d-2)\,\ell^2\,\Upsilon^{\frac{2}{d-1}}}\right)
\end{equation}
Unlike the asymptotically flat solutions studied in the previous subsection, the heat capacity may be positive for a large enough BH. The relation between the heat capacity and the size of the BH is shown in figure \ref{fig:AdSSchwarzschildHeatCapacity}. The condition $C_{D}>0$ is satisfied if the horizon radius is larger than a critical size set by the dimension and the AdS length scale
\begin{equation}\label{CriticalSize}
  r_{crit} = \sqrt{\frac{d-2}{d}}\,\ell~.
\end{equation}
This corresponds to a critical periodicity for the Euclidean time given by
\begin{equation}\label{}
  \beta_{crit} = \frac{\sqrt{d(d-2)}}{2 \pi \ell} ~.
\end{equation}
The possibility of a phase transition, as studied by Hawking and Page \cite{Hawking:1982dh}, is encountered when $r_h$ drops below the critical value given by \eqref{CriticalSize}. This is consistent with our general discussion in subsection \ref{sec:Stability}.

\subsection{Reduction of BTZ}
\label{sec:BTZ}

Gravity with a negative cosmological constant in 3 dimensions admits a solution known as the BTZ BH \cite{Banados:1992wn}. 
If $\Lambda=-1/\ell^2$ then (Euclidean) line element of the BTZ BH is
\eq{
ds^2_{BTZ}=\killing(r)d\tau^2+\frac{1}{\killing(r)}dr^2+r^2(d\phi-\frac{4G_3J}{r^2}dt)^2
}{eq:BTZ1}
where
\eq{
\killing(r)=\frac{r^2}{\ell^2} - 8G_3\mass + \frac{16G_3^2J^2}{r^2}\,.
}{eq:BTZ2}
If the inequality $|J|\leq M\ell$ holds then two horizons exist,
\eq{
r_\pm =\ell \sqrt{4G_3M} \, \sqrt{1\pm\sqrt{1-\frac{J^2}{M^2\ell^2}}}\,.
}{eq:BTZ2.5}
The normalization chosen in the rest of this work would require $G_3=1/4$, but we set $G_3=1/8$ to obtain the same normalization of the action used in \cite{Banados:1992wn,Achucarro:1993fd}. Some useful formulas are
\eq{
\frac{r_+^2+r_-^2}{\ell^2}=M\,,\qquad \frac{2r_+r_-}{\ell}= J\,,\qquad \frac{r_+^2-r_-^2}{\ell^2}=M\sqrt{1-\frac{J^2}{M^2\ell^2}}\,.
}{eq:BTZ2.6}
Achucarro and Ortiz performed a Kaluza-Klein reduction
\eq{
ds^2_{BTZ}=g_{\al\be}dx^\al dx^\be+X^2\left(d\phi+A_\al dx^\al\right)^2
}{eq:BTZ3}
to two dimensions. The dilaton field is the surface radius,
$X=r$,
and the two-dimensional line element $g_{\al\be}dx^\al dx^\be$ corresponds to the $\tau r$-part of \eqref{eq:BTZ1}. They integrated out the gauge field $A_\al$ and obtained an action of the type \eqref{Action} with \cite{Achucarro:1993fd}
\eq{
U^{AO}=0\,,\qquad V^{AO}=-\frac{X}{\ell^2} + \frac{J^2}{4X^3}\,.
}{eq:AO}
We point out an important deficiency of \eqref{eq:AO}: the angular momentum $J$ enters here as a parameter in the action rather than emerging as a constant of motion. Consequently, the thermodynamics is not reproduced correctly from the two-dimensional model \eqref{Action} with the potentials \eqref{eq:AO}. But there is a simple solution to this problem: we have demonstrated in subsection \ref{sec:Maxwell} how to treat charged BHs and noted that upon introducing Maxwell fields a parameter in the action may be converted into a constant of motion, namely a conserved charge. In fact, the Kaluza-Klein Ansatz \eqref{eq:BTZ3} already contains such a Maxwell field and all one has to do is to refrain from integrating it out. Therefore, we study an action \eqref{actionmax} [multiplied by 2 to get the normalizations above] with the functions
\begin{align} \label{BTZU}
 U^{BTZ}(X) &= 0\,, \\ \label{BTZV}
 V^{BTZ}(X) &= - \frac{X}{\ell^2}\,, \\ \label{BTZf}
 f^{BTZ}(X) &= -\frac{1}{8}\,X^3\,.
\end{align}
The potentials \eqref{BTZU} and \eqref{BTZV} are consistent with \eqref{ReducedU} and \eqref{ReducedV}, respectively, for $d+1=3$. The Maxwell field with coupling function \eqref{BTZf} arises because we are treating a spinning BH rather than a spherically symmetric one. The conserved $U(1)$ charge therefore has to be identified with the angular momentum, $q\propto J$. The proportionality factor depends on the conventions. With our choices the correct relation is $q=J/2$. For further calculations we need the functions
\eq{
Q^{BTZ}(X)=0\,,\qquad w^{BTZ}(X)=\frac{X^2}{\ell^2}\,,\qquad h^{BTZ}(X)=\frac{4}{X^2}
}{eq:BTZ4}
and note that the ground state solution $\mass=J=0$ is pure AdS$_3$, $\killing_0=X^2/\ell^2$.

We show now that this effective two-dimensional description leads to the correct thermodynamics for the BTZ BH. To this end we apply the general results of subsection \ref{sec:Maxwell} to the specific choice \eqref{eq:BTZ4} recalling that $q=J/2$. The square of the Killing norm
\eq{
\killing(X)=\frac{X^2}{\ell^2}-M+\frac{J^2}{4X^2}
}{eq:BTZ5}
indeed is equivalent to \eqref{eq:BTZ2} for $G_3=1/8$ and $X=r$. This together with the gauge potential
\eq{
A_r=0\,,\qquad A_\tau=- \frac{J}{2X^2}\,,
}{eq:BTZ6}
plugged into \eqref{eq:BTZ3} allows to recover the BTZ line-element \eqref{eq:BTZ1}. With the proper angular momentum [cf.~\eqref{eq:max6}]
\eq{
\Omega_c:= \Phi_c=\frac{J}{2\sqrt{\killing_c}} \left(\frac{1}{X_h^2}-\frac{1}{X^2_c}\right)
}{eq:BTZ12}
the thermodynamic potential \eqref{eq:Y} is given by
\eq{
Y(T_c,X_c,\Omega_c)=-T_cS + E_c - \Omega_c J \,.
}{eq:BTZ7}
Entropy is determined by the value of the dilaton at the outer horizon, $X_h=r_+$. Taking into account that the normalization of Newton's constant differs in this subsection from the one used in the rest of this work by a factor of 2 we get from \eqref{Entropy2}
\eq{
S=4\pi r_+ \,.
}{eq:BTZ8}
The result \eqref{eq:BTZ8} coincides with the entropy of the BTZ calculated by counting of microstates, cf.~e.g.~\cite{Carlip:1998qw}. The inverse Euclidean periodicity determines temperature,
\eq{
T=\frac{r_+^2-r_-^2}{2\pi r_+\ell^2} = \frac{2M}{S}\sqrt{1-\frac{J^2}{M^2\ell^2}}\,.
}{eq:BTZ9}
This coincides with the result for the temperature of the BTZ BH \cite{Banados:1992wn}. Obviously, in the limit $|J|/(M\ell) \to 1$ temperature vanishes, which is consistent with the fact that for $|J|=M\ell$ the BTZ BH becomes extremal. If we define the quantity $\Omega$ as
\eq{
\Omega:=\lim_{X_c\to\infty}\sqrt{\killing_c}\Omega_c=\frac{J}{2r_+^2}
}{eq:BTZ10}
then we obtain the correct first law relating the conserved charges,
\eq{
dM=TdS+\Omega dJ\,.
}{eq:BTZ11}
The integrated version
\eq{
2M=TS+2\Omega J
}{eq:BTZ16}
shows that extensitivity properties differ slightly from standard thermodynamics. However, they are as expected for a BH in AdS: if the dilaton (and thus entropy) is extensive then the square-root of the mass $\mass$ is extensive, consistently with the extensitivity properties in table \ref{tab:1}. This explains the factor of 2 on the left hand side of \eqref{eq:BTZ16}. The factor 2 on the right hand side is the standard result for spinning BHs. With the results above, the expression for internal energy [note again the factor of 2 as compared to \eqref{eq:td10} because of the different normalization of the action],
\eq{
E_c(T_c,X_c,J)=2\sqrt{\killing_0}-2\sqrt{\killing_c}=\frac{2X_c}{\ell}\left(1-\sqrt{1-\frac{\mass\ell^2}{X_c^2}+\frac{J^2\ell^2}{4X_c^4}}\right)\,,
}{eq:BTZ14}
and surface pressure
\eq{
\psi_c=\frac{E_c}{\ell\sqrt{\killing_c}}-\frac{J^2}{2X_c^3\sqrt{\killing_c}}
}{eq:BTZ15}
lead to the quasilocal form of the first law
\eq{
dE_c=T_c dS + \Omega_c dJ - \psi_c dX_c ~.
}{eq:BTZ13}
All our results are compatible with \cite{Brown:1994gs} (cf.~also \cite{Zaslavsky:1997ha}). 
This shows that the BTZ BH thermodynamics is compatible with a Kaluza-Klein reduction to two dimensions. A generalization to a charged BTZ BH is straightforward and requires the addition of another Maxwell field with a linear coupling to the dilaton. A different generalization involves the coupling to a gravitational Chern-Simons term \cite{Deser:1982vy}, 
which yields upon dimensional reduction \cite{Guralnik:2003we} a model of type \eqref{actionmax} \cite{Grumiller:2003ad} whose entropy recently has been derived in \cite{Sahoo:2006vz} using Wald's Noether charge technique \cite{Wald:1993nt}.

\section{Discussion}
\label{sec:Discussion}

In this paper we have studied black hole thermodynamics for arbitrary models of dilaton gravity in two dimensions. The analysis of these models was complicated by the fact that the path integral defined with respect to the action \eqref{Action} does not have a sensible semi-classical limit. We resolved this problem by constructing an improved action,
\begin{multline}\label{ActionConclusion}
  \Gamma = - \frac{1}{16\pi G_2}\,\int_{\MM} \nts \nts d^{\,2}x \,\sqrt{g}\, \left[ X\,R - U(X)\,
        \left(\nabla X\right)^2 - 2 \, V(X) \right] \\
- \frac{1}{8\pi G_2}\, \int_{\dM} \bns dx \, \sqrt{\gamma}\,X\,K
+ \frac{1}{8\pi G_2}\, \int_{\dM} \bns dx \,\sqrt{\gamma}  \,
    \sqrt{w(X) \, e^{-Q(X)}} \,,
\end{multline}
generalizing the techniques first used in \cite{Davis:2004xi} to obtain the boundary counterterm \eqref{ICT}. The counterterm is unique up to a shift ambiguity\footnote{In addition there is a sign ambiguity which is trivially fixed by demanding finiteness of the improved on-shell action.} which we fixed by demanding that it be independent under changes of scale associated with constant shifts in $Q(X)$. Thus, one should consider \eqref{ActionConclusion}, instead of \eqref{Action}, as `the generic dilaton gravity action'. Following the approach of York \cite{York:1986it}, we proceeded to define the canonical partition function for a black hole enclosed in the `cavity' $X \leq X_c$. The Helmholtz free energy \eqref{FreeEnergy} for this system is related to the improved action by $F_c = \beta_{c}^{-1} \, \Gamma_c$, where the proper inverse temperature $\beta_{c}^{-1}$ is fixed at the cavity wall $X_c$ by an external thermal reservoir. All of the results on thermodynamics in section \ref{sec:Thermo} follow from this identification. After deriving various thermodynamical quantities, we generalized our results to include two dimensional Maxwell fields and their associated conserved charges. We then examined the conditions under which the system is thermodynamically and mechanically stable, and described processes which might destabilize the system. Our results were then applied to a wide range of models in sections \ref{sec:2DExamples}-\ref{sec:HigherDimensionalExamples}. This includes backgrounds that appear in two-dimensional solutions of string theory (section \ref{sec:StringExamples}). Finally, we discovered that our results can be used to describe the thermodynamics of certain higher dimensional black holes (section \ref{sec:HigherDimensionalExamples}).

An important aspect of our analysis is universality. For instance, we confirmed that the entropy \eqref{Entropy2} for a dilaton gravity black hole in two dimensions is essentially given by $X_h$, the dilaton evaluated at the horizon. This result is independent of the details that define a particular model. If we define the `area' of the horizon via the $d \to 1$ limit of a $d$ dimensional sphere and restore Newton's constant, then this result can be expressed as
\begin{equation}\label{LastEntropy}
    S = \frac{A_h}{4\,G_{\rm eff}} ~.
\end{equation}
The effective Newton's constant is defined as $G_{\rm eff} = G_2 / X_h$. Since $G_2$ is dimensionless there is no notion of Planck length in two dimensions. However, we can think of $G_{\rm eff}$ as defining an effective resolution at the horizon, in the sense that decreasing $G_{\rm eff}$ increases the number of bits that can be stored by the black hole. Then \eqref{LastEntropy} can be interpreted in much the same way as the usual relationship between entropy and area in higher dimensions. In section \ref{sec:Energy} we obtained a simple expression for the (proper) internal energy of the system
\begin{equation}\label{LastEnergy}
  E_c = e^{-Q(D_c)} \, \left(\sqrt{\xi_0} - \sqrt{\xi_c}\right) ~.
\end{equation}
This agrees with the calculation of the energy using the quasilocal stress tensor defined by Brown and York \cite{Brown:1991gb}. We were able to show that every dilaton gravity black hole satisfies a quasilocal form of the first law of black hole thermodynamics
\begin{equation}
    dE_c = T_c \, dS - \psi_c \, d D_c + \Phi_{c} \, d q ~.
\label{eq:con1}
\end{equation}
This expression holds at arbitrary values of $X_c$, incorporates the dilaton charge $D_c$ and its chemical potential $\psi_c$, allows for Maxwell fields with proper electrostatic potentials $\Phi_c$ and conserved charges $q$, and accounts for the (non-linear) effects of gravitational binding energy present in $E_c$. The canonical first law \eqref{eq:con1} contains considerably more information than the microcanonical $dM = T \, dS$, which becomes a triviality if expressed as $dw_h=w_h'dX_h$.

Despite our best attempts to provide a comprehensive study, there are several directions that we have not had a chance to explore. A number of `working assumptions' were made throughout the paper and it is important to understand which ones can be relaxed. An example is our decision to focus on models where $w \to \infty$ as $X \to \infty$. This is a reasonable assumption, as it seems to be the most common behavior among dilaton gravity models. But there are certainly models for which it does not apply. For instance, one might consider models where $w$ approaches a constant, or $w \to - \infty$, as $X \to \infty$. Our approach should still be applicable, though many of the considerations in section \ref{sec:Thermo} must be modified.  In the first case, where $w$ approaches a constant, we can make a constant shift so that $w \to 0$ as $X \to \infty$. Then one can exploit the duality described in \cite{Grumiller:2006xz}, which preserves the classical solution for the metric but replaces $w$ with $\tilde{w}=1/w$. This implies $\tilde{w} \to \infty$ as $X \to \infty$, which can be analyzed under our initial assumptions. The second case, $w \to -\infty$ as $X \to \infty$, is perhaps more interesting as it includes the dimensional reduction of the static patch of de Sitter space. We now have to consider the region $X \leq X_{dS}$, where $X_{dS}$ corresponds to the observer-dependent cosmological horizon in the static patch. Although we have not performed this analysis, we expect that our approach should reproduce the standard results for the thermodynamics of cosmological horizons. An additional complication emerges when one considers a second horizon, $X_h < X_{dS}$, associated with a black hole. The analysis of that system is beyond the scope of this paper.

It would be interesting to study the impact of the boundary counterterm in \eqref{ActionConclusion} on the exact path integral quantization procedure in the first order formulation of dilaton gravity, cf.~e.g.~\cite{Kummer:1996hy,Grumiller:2002nm}. 
While the bulk term and the Gibbons-Hawking-York boundary term look quite different from \eqref{Action} in the first order formulation, we expect that the boundary counterterm \eqref{ICT} can be translated easily to its first order counterpart, because it depends solely on the dilaton field and the induced volume form at the boundary.

Another idea that deserves to be studied in more detail was mentioned at the end of subsection \ref{sec:Maxwell}. Consider a dilaton gravity model with a potential of the form \eqref{eq:max8}. The same model can be obtained by integrating out a number of Maxwell fields whose coupling functions $f(X)$ are related to terms in the potential. This means that various coupling constants in the potential $V(X)$ can be converted to conserved $U(1)$ charges associated with Maxwell fields. This technique allowed us to study the thermodynamics of the BTZ BH in subsection \ref{sec:BTZ}. The first law for the rotating, three dimensional solution was recovered by `integrating in' a gauge field that Achucarro and Ortiz had integrated out in their toroidal reduction \cite{Achucarro:1993fd}. It would be interesting to study the extent to which the thermodynamics of higher dimensional solutions can be reproduced in this manner.

The most pressing generalization of our work involves coupling two dimensional dilaton gravity to matter, such as a scalar field. This is important for several reasons. As mentioned in the introduction, we have assumed that these models will be coupled to some form of propagating matter, and scalar fields are a prime candidate. We have also considered backgrounds that arise in two dimensional solutions of string theory, and any further analysis of these models should try to incorporate the tachyon. In general, coupling additional fields to the dilaton gravity model will introduce complications at the level of the Hamilton-Jacobi equation. However, even if this equation can no longer be solved exactly, it can still be addressed perturbatively. This approach has been used successfully for gravity coupled to a scalar field in higher dimensions \cite{deBoer:1999xf, Martelli:2002sp, Larsen:2004kf, Batrachenko:2004fd}. We hope to address this in a future publication.

\acknowledgments

The authors would like to thank Josh Davis, Roman Jackiw, Antal Jevicki, Mark Spradlin, Alessandro Torrielli and Dima Vassilevich for useful discussions. In addition DG would like to thank Jan {\AA}man, Ingemar Bengtsson and Narit Pidokrajt for the hospitality at Stockholm University and for their interest. The work of DG is supported in part by funds provided by the U.S. Department of Energy (DoE) under the cooperative research agreement DEFG02-05ER41360. DG has been supported by the Marie Curie Fellowship MC-OIF 021421 of the European Commission under the Sixth EU Framework Programme for Research and Technological Development (FP6). The research of RM was supported in part by the DoE through Grant DE-FG02-91ER 40688, Task A (Brown University).

\appendix

\section{Conventions, Dimensions, and a Dilaton Gravity Bestiary}
\label{sec:Conventions}

Let $\MM$ be a two-dimensional manifold with boundary $\dM$. We consider a metric $g_{\mu\nu}$ on $\MM$ that induces a metric $\gamma_{ab}$ on $\dM$. The (one-dimensional) boundary indices are useful for keeping track of factors of $\gamma$ and its inverse.

The Riemann tensor on $(\MM,g)$ and its contractions are defined so that spheres have positive curvature and hyperbolic spaces have negative curvature
\begin{gather} \label{RiemannDef}
  R^{\lambda}_{\dub \mu \rho \nu} = \partial_{\rho} \Gamma^{\lambda}_{\mu\nu} + \Gamma^{\kappa}_{\mu\nu}
     \Gamma^{\lambda}_{\kappa \rho} - \left( \nu \leftrightarrow \rho \right) \\ \label{RicciDef}
  R_{\mu\nu} =  R^{\lambda}_{\dub \mu \lambda \nu} ~.
\end{gather}
The extrinsic curvature associated with the embedding of $(\dM,\gamma)$ in $(\MM,g)$ is given by
\begin{equation}
   K_{ab} = \frac{1}{2} \, n^{\mu} \partial_{\mu} \gamma_{ab}
\end{equation}
where $n^{\mu}$ is an outward-pointing unit vector normal to $\dM$. The trace of the extrinsic curvature is $K = \gamma^{ab} K_{ab}$. Because the boundary is one-dimensional, it follows that
\begin{equation}
   K_{ab} - \gamma_{ab} \, K = 0 ~.
\end{equation}
This implies that $\pi^{ab}$ would vanish if not for the coupling between $X$ and the Ricci scalar in \eqref{Action}.

Throughout the paper we work with Euclidean metrics of the form
\begin{equation}\label{metric2}
  ds^2 = \xi(r) \, d\tau^2 + \frac{1}{\xi(r)} \, dr^2 ~.
\end{equation}
Calculating the action and its variation requires the Ricci scalar and the extrinsic curvature of the boundary. The Ricci scalar is given by
\begin{equation}\label{eq:R}
 R = - \frac{\partial^{2}\xi}{\partial \,r^{\,2}} = -e^{-Q}\left[w''+Uw'+U'(w-2M)\right]~.
\end{equation}
The last expression, which follows from \eqref{XPrimeDef} and \eqref{xiDef}, is coordinate independent and relates the Ricci scalar to the dilaton field. For models obeying the Minkowskian ground state condition \eqref{eq:MGS1} it simplifies to $R=2M({w'}^2/w-w'')$. Boundary quantities are obtained from the $X \to \infty$ limit of a constant $X$ surface, as described in subsection \ref{sec:BlackHoles}. The induced metric on the surface is $\gamma_{ab} = \xi(X)$, and the trace of its extrinsic curvature is
\begin{equation}\label{eq:K}
  K = \frac{1}{2 \, \sqrt{\xi}}\,\frac{\partial \xi}{\partial r} = e^{-Q/2}\,\frac{(w-2M)U+w'}{2\sqrt{w-2M}}~.
\end{equation}
 For models obeying the Minkowskian ground state condition \eqref{eq:MGS1} it simplifies to $K=Mw'/(w\sqrt{\killing})$.
The Euler characteristic for the space-time is related to \eqref{eq:R} and \eqref{eq:K} by
\begin{equation}
   4\pi \, \chi = \int_{\MM} \nts \nts d^{\,2}x \sqrt{g}\,R + 2 \, \int_{\dM}\bns dx \sqrt{\gamma} \, K ~.
\end{equation}
One can verify that the BH solutions of discussed in \ref{sec:BlackHoles} have the topology of a disk, $\chi = 1$.

The dimensions of various quantities are assigned conveniently, but at times differently, in the examples we consider. Here we clarify the role of physical dimensions, how to fix them, and how to rescale them. Unless stated otherwise we always work in natural units $\hbar=c=k_{\rm B}=1$ and choose the dilaton field $X$ to be dimensionless. This means that the potential $U(X)$ is dimensionless, as well. If the coordinates are taken to have the natural dimension of length then the relation \eqref{XPrimeDef} implies that $e^{Q}$ must also have dimensions of length. This can be accomplished by defining $e^{Q} = \lambda \exp{\int^{X} d\tilde{X} U(\tilde{X})}$, where $\lambda$ is a dimensionful constant with units of length that includes the factor of $e^{Q_0}$. In that case, both $w$ and $\mass$ must have units of inverse-length, so that \eqref{xiDef} is dimensionless. Similarly, coordinates with units of length imply that the function V has dimensions of one over length squared. Units of length can then be changed by appropriate rescalings of the dimensionful constants in $V$, as well as $\lambda$ and $M$. Alternatively, if the coordinates are taken to be dimensionless then $U$, $e^{Q}$, $V$, $w$, and $M$ are all dimensionless, as well.

After these general observations we focus on the choices of relevance for our work. With one exception (mentioned below) we demand that the dilaton field $X$ be dimensionless. In some of our examples we do not exhibit the dimensionful constants, for simplicity. Instead we treat all quantities as dimensionless, including the line-element, and restore appropriate physical dimensions when necessary. This applies to the discussion of the $ab$-family in subsection \eqref{sec:abFamily}, and to the discussion of the BTZ BH in subsection \eqref{sec:BTZ} if $X,\ell,J,\mass$ are considered to be dimensionless. In other examples we keep track of the physical dimensions for easier comparison with the literature. This applies to sections \ref{sec:StringExamples}, \ref{sec:ReducedSchwarzschild}, and \ref{sec:AAdS}.  Alternatively, in subsection \ref{sec:BTZ} the quantities $X,\ell,J$ can be thought of as having dimensions of length, with $M$ dimensionless.

\TABLE{
\renewcommand{\arraystretch}{1.5}
\begin{tabular}{ | l || c | c | c | c || c | } \hline
  Model \raisebox{-6pt}[14pt]{} & $U(X)$ & $V(X)$ & $e^{Q(X)}$ & $w(X)$ & Reference \\ \hline \hline
  Schwarzschild & $- \frac{1}{2 X}$ & $-\frac{1}{2G_4}$ & $\sqrt{\frac{G_4}{2X}}$ & $\sqrt{\frac{2X}{G_4}}$ & \cite{Schwarzschild:1916uq} \\ 
  Jackiw-Teitelboim & $0$ & $-\Lambda X$ & $1$ & $\Lambda X^{2}$ & \cite{JT} \\ 
  Witten BH & $-\frac{1}{X}$ & $-\frac{\lambda^2}{2} \, X$ & $\frac{1}{\lambda X}$ & $\lambda X$ & \cite{Witten:1991yr} \\ 
  CGHS & $0$ & $-\frac{\lambda}{2}$ & $1$ & $\lambda X$ & \cite{Callan:1992rs} \\ 
  (A)dS$_2$ Ground State & $-\frac{a}{X}$ & $-\frac{1}{2} \, B X$ & $\frac{1}{X^{a}}$ & $\frac{B}{2-a}\,X^{2-a}$ & \cite{Lemos:1994py} \\ 
  Rindler Ground State & $-\frac{a}{X}$ & $-\frac{1}{2} \, B X^{a}$ & $\frac{1}{X^{a}}$ & $B X$ & \cite{Fabbri:1996bz} \\ 
  BH Attractor & $0$ & $-\frac{B}{2 X}$ & $1$ & $B \ln{X}$ & \cite{Grumiller:2003hq} \\ 
  $ab$-Family & $-\frac{a}{X}$ & $-\frac{B}{2} \, X^{a+b}$ & $\frac{1}{X^{a}}$ & $\frac{B}{b+1}\,X^{b+1}$ & \cite{Katanaev:1997ni} \\ 
  Liouville Gravity & $a$ & $b e^{\alpha X}$ & $e^{a X}$ & $-\frac{2b}{a+\alpha} \, e^{(a+\alpha) X}$ & \cite{Nakayama:2004vk} \\ 
  Exact String BH & \cite{Grumiller:2005sq} & \cite{Grumiller:2005sq} & \eqref{eq:EBH6} & \eqref{eq:EBH6}, \eqref{eq:EBH2} & \cite{Dijkgraaf:1992ba} \\
  Schwarzschild-(A)dS \raisebox{-8pt}[14pt]{}  & $-\frac{1}{2 X}$ & $-\frac{1}{2G_4} - \frac{3}{\ell^2}\,X$ & $\sqrt{\frac{G_4}{2X}}$ & $\sqrt{\frac{2X}{G_4}} \big( 1+  \frac{2G_4}{\ell^2}\,X\big)$ & \cite{Hawking:1982dh} \\ 
  Katanaev-Volovich & $\alpha$ & $\beta X^2 - \lambda$ & $e^{\alpha X}$ & $\frac{2}{\alpha} \, e^{\alpha X} \, \left( \lambda -\beta X^2 + \raisebox{12pt}{}\right.$ & \cite{Katanaev:1986wk} \\
   \raisebox{-8pt}[14pt]{} &   &   &   & $\left. + 2 \frac{\beta}{\alpha}\,X - 2 \frac{\beta}{\alpha^2}\right)$ &   \\ 
  KK Reduced CS & $0$ & $\frac{1}{2} \, X \left(c-X^2\right)$ & $1$ & $\frac{1}{4} \left( c-X^2 \right)^2 - \frac{1}{4}\,c^2$ & \cite{Guralnik:2003we,Grumiller:2003ad} \\ 
  \hline
\end{tabular}
\caption{Summary of Dilaton Gravity models, adapted from \cite{Grumiller:2006rc}.}
\label{ModelTable1}
}

\TABLE{
\renewcommand{\arraystretch}{1.5}
\begin{tabular}{ | l || c | c | c | c | c | c || c | } \hline
  Model \raisebox{-6pt}[14pt]{} & $U(X)$ & $V(X)$ & $f(X)$ & $e^{Q(X)}$ & $w(X)$ & $h(X)$ & Reference \\ \hline \hline
Reissner-Nordstr\"{o}m & $-\frac{1}{2 X}$ & $-\frac{1}{2G_4}$ & $X$ & $\sqrt{\frac{G_4}{2X}}$ & $\sqrt{\frac{2X}{G_4}}$ & $-\sqrt{\frac{2G_4}{X}}$ & \cite{Reissner:1916} 
\\
 Achucarro-Ortiz & 0 & $-\frac{X}{\ell^2}$ & $-\frac{X^3}{8}$ & $1$ & $\frac{X^2}{\ell^2}$ & $\frac{4}{X^2}$ & \cite{Achucarro:1993fd} \\
 2D Type 0A/0B  & $-\frac{1}{X}$ & $-\frac{\lambda^2}{2} X$ & $\pi\alpha^\prime$ & $\frac{1}{\lambda X}$ & $\lambda X$ & $\frac{\ln{X}}{\lambda\pi\alpha^\prime}$ & \cite{Berkovits:2001tg} \\ \hline
\end{tabular}
\caption{Dilaton Gravity models with Maxwell fields.}
\label{ModelTable2}
}

The analysis of sections \ref{sec:SPI} and \ref{sec:Thermo} is carried out for a general model of two-dimensional dilaton gravity. As a convenience, the properties of several models are summarized in tables \ref{ModelTable1} and \ref{ModelTable2}. Table \ref{ModelTable1} lists models that contain only a metric and a dilaton, while table \ref{ModelTable2} lists models that also contain Maxwell fields. The last three models in table \ref{ModelTable1} can be converted into models belonging to table \ref{ModelTable2} as explained at the end of subsection \ref{sec:Maxwell}. In most entries in these tables we have chosen the scaling and shift ambiguities inherent to $Q$ and $w$ as to achieve coincidence with the choices in the main text. In some brief examples in our paper the four-dimensional Newton constant is fixed as $G_4=1/2$, thus simplifying the first entries in both tables, as well as the Schwarzschild-(A)dS entry in table \ref{ModelTable1}. In subsection \ref{sec:abFamily} we set $B=b+1$, which simplifies the corresponding entry '$ab$-Family' in table \ref{ModelTable1}.

\section{Weyl Rescaled Metric}
\label{sec:NewVariables}

In subsections \ref{sec:Variation} and \ref{sec:FinalAction} we employ a change of variables that involves a Weyl rescaling of the metric. Then, in subsection \ref{sec:Conformal}, we consider the effect of a conformal transformation on a model. In this appendix we work out some of the relevant details.

Consider a new metric related to $g_{\mu\nu}$ by
\begin{equation}\label{newmetric}
g_{\mu\nu} = e^{2\sigma(X)}\,\hg_{\mu\nu}
\end{equation}
where $\sigma(X)$ is an arbitrary function of the dilaton. If we express the action \eqref{Action} in terms of this new metric we find
\begin{equation}\label{NewVariableAction}
   I[X,g_{\mu\nu}] = - \frac{1}{2} \, \int_{\MM} \nts d^{\,2}x \sqrt{\hg}\,\left[X \hR - \hat{U}(X) (\hnab X )^2 -2 \, \hV(X)\right] - \int_{\dM} \bns dx \sqrt{\hgam} X \hK\,.
\end{equation}
This has the same functional form as \eqref{Action}, with dilaton potentials $\hU$ and $\hV$ that are related to $U$ and $V$ by
\begin{equation}
    \hU(X) = U(X) - 2 \sigma'(X) \quad \quad \hV(X) = e^{2\sigma(X)} \, V(X) ~.
\label{eq:apB7}
\end{equation}
The functions $\hQ(X)$ and $\hw(X)$ transform according to
\begin{equation}
    \hQ(X) = Q(X) - 2\,\sigma(X) \quad \quad \hw(X) = w(X) ~.
\label{eq:apB13}
\end{equation}
A change of variables with $2\sigma(X) = Q(X)$ leads to vanishing $\hat{U}$. This transformation is common in the dilaton gravity literature because it simplifies many calculations. Notice that $w(X)$ is invariant under \eqref{newmetric}. Referring to \eqref{DivergentAction}, which depends only on $w(X)$, $X$, and the parameter $M$, we confirm that the on-shell action is invariant. This is as it should be: the value of the action cannot depend on the variables we use to evaluate it.

Now we want to examine the variation of the action using the new metric variable. In terms of $\hg_{\mu\nu}$ and $\sigma$ a general variation of $g_{\mu\nu}$ takes the form
\begin{equation}
   \delta g_{\mu\nu} = e^{2\sigma}\,\delta\hg_{\mu\nu} + 2 \,e^{2\sigma}\hg_{\mu\nu} \sigma' \delta X ~.
\end{equation}
The change in the action \eqref{ActionVariation} due to a small variation of the fields can be written as
\begin{equation}\label{NewVariation}
 \delta I[X, g_{\mu\nu}] = \int_{\MM} \nts d^{\,2}x \sqrt{\hg}\,\left[\hat{\EE}^{\mu\nu}
    \delta \hg_{\mu\nu} + \hat{\EE}_{X} \, \delta X \right]
    + \int_{\dM} \bns dx \sqrt{\hgam} \, \left[ \hat{\pi}^{ab} \delta \hgam_{ab} + \hat{\pi}_{X}
    \delta X \right] ~.
\end{equation}
Both the equations of motion $\hat{\EE}_{\mu\nu}$ and $\hat{\EE}_{X}$, and the momenta $\hat{\pi}^{ab}$ and $\hat{\pi}_{X}$ have the same functional form as in the main text, with all un-hatted quantities being replaced by their hatted counterparts. The solution of the equations of motion is
\begin{align}\label{NewSolution}
   X &= X(\hat{r}) & d\hat{s}^2 &= \hxi(\hat{r}) \,d\tau^2 + \frac{1}{\hxi(\hat{r})}\,d\hat{r}^2 \\
  \partial_{\hat{r}} X  &=  e^{-\hQ(X)} & \hxi(X)  &=  e^{\hQ(X)}\,\left(w(X) - 2\,M \right) ~.
\end{align}
This is just the solution from section \ref{sec:Action}, expressed in terms of the new metric and written in a coordinate system $(\tau,\hat{r})$ with $\partial_{\hat{r}} = e^{2\sigma(X)} \partial_{r}$. Evaluating \eqref{NewVariation} for this solution gives
\begin{equation}\label{NewVariation2}
  \left. \delta I[X,g_{\mu\nu}] \raisebox{12pt}{}\right|_{\hat{\EE}=0} = \int_{\dM}\bns dx \,
  \left[ - \frac{1}{2} \, e^{-\hQ} \, \delta \hxi +  e^{-\hQ}\left( \hU \hxi - \frac{1}{2}\,\hxi'
  \right) \delta X \right]\,.
\end{equation}
We can now choose the function $\sigma(X)$ in \eqref{newmetric} to simplify the analysis of this expression. If we define
\begin{equation}
  e^{2\sigma(X)} = w(X) \, e^{Q(X)}\,,
\label{eq:apB3}
\end{equation}
then the (square of the) Killing norm for the new metric is given by
\eq{
\hxi(X) = 1 - \frac{2M}{w(X)} ~.
}{eq:apB1}
This makes the meaning of Dirichlet boundary conditions unambiguous: $\hxi$ takes the constant value $\hxi=1$ at $\dM$, and any variation $\delta \hxi$ that preserves this condition must vanish at $\dM$. The functions $\hQ$ and $\hU$ are determined as
\begin{equation}\label{NewFunctions}
 e^{\hQ(X)} = \frac{1}{w(X)} \qquad \hU = - \frac{w'(X)}{w(X)}
\end{equation}
so that \eqref{NewVariation2} yields
\begin{equation}\label{FinalNewVariation}
 \delta I_{\rm reg} = \int d\tau \, \left[ -\frac{1}{2} \, w \, \delta \hxi -
   \left(1-\frac{M}{w}\right)\,w' \, \delta X \right] ~.
\end{equation}
The subscript indicates that this quantity is evaluated using a regulator, as in section \ref{sec:Variation}. Following the arguments in the main text we ignore the $\delta X$ term and concentrate on a variation of the metric $\delta \hxi = \delta M e^{\hQ} = \delta M / w$. The corresponding change in the action is
\begin{equation}
    \lim_{X_{\rm reg} \to \infty} \delta I_{\rm reg} = \int d\tau \, \delta M \neq 0
\end{equation}
which of course agrees with \eqref{deltaI}. Changing variables to a metric that obeys standard Dirichlet conditions at $\dM$ simplifies the analysis of the improved action, as well, and is useful in establishing the result \eqref{equationwithoutlabel}.

In subsection \ref{sec:Conformal} we treat \eqref{newmetric} as a Weyl transformation, as opposed to a change of variable. Then the action \eqref{NewVariableAction} is interpreted as a functional of $X$ and $\hg_{\mu\nu}$. This gives a new model of type \eqref{Action} with potentials given by \eqref{eq:apB7}.

\section{Proof of Gauge-Independence}
\label{sec:GaugeIndependence}

In this appendix we show that the main results of section \ref{sec:SPI} are independent of the choice of gauge  \eqref{metric}. We begin by establishing two important properties of solutions of \eqref{MetricEOM} and \eqref{XEOM}. Consider the vector $k^{\mu}$ defined by
\begin{equation}\label{KillingVector}
  k^{\mu} = \epsilon^{\mu\nu} \, e^{Q(X)} \, \nabla_{\nu} X
\end{equation}
where $e^{Q(X)}$ was given in \eqref{QDef}. The Lie derivative of the dilaton along $k^{\mu}$ vanishes by construction
\begin{equation}\label{ConstantX}
  \LL_{k} X = e^{Q(X)} \epsilon^{\mu\nu} \, \partial_{\mu} X \partial_{\nu} X = 0 ~.
\end{equation}
The same result also applies to the metric. It follows from the equation of motion \eqref{MetricEOM} that the vector $k^{\mu}$ satisfies Killing's equation
\begin{equation}\label{}
  \LL_{k} g_{\mu\nu} = \nabla_{\mu} k_{\nu} + \nabla_{\nu} k_{\mu} = 0 ~.
\end{equation}
Therefore, solutions of \eqref{MetricEOM} and \eqref{XEOM} always possess at least one Killing vector whose orbits are isosurfaces of the dilaton.

The properties of $k^{\mu}$ allow us to construct a coordinate system $(\tau,r)$ where the Killing vector is $k^{\mu} \partial_{\mu} = \partial_{\tau}$, and the metric and dilaton take the form
\begin{equation}\label{Decomposition}
   ds^{2} = N(r)^2 \, dr^2 + \xi(r)\,\left( d\tau + N^{\tau}(r)\,dr \right)^2   \quad \quad X=X(r) ~.
\end{equation}
The functions $N$ and $N^{\tau}$ that appear in the metric are the lapse function and shift `vector', respectively, associated with evolution in the direction orthogonal to $k^{\mu}$. Normally, the lapse and shift both appear in the action as Lagrange multipliers. In this case, the fact that all fields are constant along the direction singled out by the Killing vector implies that the action is independent of $N^{\tau}(r)$. Furthermore, the coefficient of $\delta N^{\tau}$ in the variation of the action vanishes identically. This means that we can freely set $N^{\tau}=0$ in \eqref{Decomposition} with no loss of generality. The metric then simplifies to
\begin{equation}\label{}
  ds^2 =  N(r)^2 \, dr^2 + \xi(r)\,d\tau^2 ~.
\end{equation}
One can substitute this form of the metric into the action and recover all of the results of a fully covariant analysis. From the point of view of the action, the functions $\xi(r)$ and $N(r)$ must be treated as independent fields. However, we are always free to present a solution in a coordinate system where $N(r)^2 = \xi(r)^{-1}$, as in \eqref{metric}.

Now we want to check whether or not the conclusions of subsection \ref{sec:Variation} are sensitive to the choice of gauge. Integrating the equations of motion for an arbitrary lapse function, we find
\begin{equation}\label{GeneralSolution}
  \begin{aligned}
    \partial_r X &\, = N(r)\sqrt{\xi(X)}\,e^{-Q(X)} \\
    \xi(X) &\, = w(X) \,e^{Q(X)} \, \left(1-\frac{2\,M}{w(X)}\right) ~.
  \end{aligned}
\end{equation}
The only difference between these expressions and the solution \eqref{XPrimeDef} and \eqref{xiDef} is an overall factor that appears in $\partial_r X$. Notably, we obtain the same expression for $\xi(X)$, concurrent with its coordinate independent interpretation as square of the Killing norm. The change in the action due to a variation in the metric, with the bulk terms set to zero, is now given by
\begin{equation}\label{Variation2}
  \delta I_{\rm reg} = -\frac{1}{2} \, \int d\tau \, \frac{\partial_r X}{N(r)\sqrt{\xi(r)}}\, \delta \xi(r) ~.
\end{equation}
The subscript is a reminder that the integrand is evaluated using a regulator. This result can be obtained either by evaluating the covariant expression \eqref{ActionVariation} for the solution \eqref{GeneralSolution}, or by expressing the action \eqref{Action} in terms of $\xi(r)$ and $N(r)$ and then varying the fields. It is clear from \eqref{GeneralSolution} that the explicit factor of $N(r)$ in \eqref{Variation2} cancels against the factor in $\partial_r X$, and we recover the result \eqref{deltaI}. Therefore, the  conclusions of subsection \ref{sec:Variation} are independent of the choice of gauge. This exercise can be repeated for the variation of the action $\Gamma$ from \eqref{GammxiDef2}. Regardless of the choice of gauge, $\delta \Gamma = 0$ for all variations of the fields that preserve the boundary conditions.

\section{Formulas}
\label{sec:Formulas}

The variations of $X_h$ and $T$ can be expressed entirely in terms of $d \mass$, using the definitions of the horizon \eqref{Horizon} and $T$ \eqref{T}
\begin{equation}\label{dX_and_dT}
  d X_h = \frac{1}{2  \pi  T}\,d \mass \quad \quad \quad  d T = \frac{1}{2 \pi T}\,\frac{\partial T}{\partial X_h}
  \,d \mass ~.
\end{equation}
Similarly, the definitions of $\killing$ \eqref{xiDef}, $Q$ \eqref{QDef}, and $w$ \eqref{wDef} give
\begin{equation}\label{dAc}
  d\killing_c = - 2\,e^{Q_c}\,d\mass + \left( U_c \, \killing_c + \, e^{Q_c}\,w'_c \right) \, dD_c ~.
\end{equation}
When $D_c$ is held fixed the variation of $\killing_c$ reduces to
\begin{equation}\label{dA_D}
  d\killing_c \Big|_{D_c} = -2 \, e^{Q_c} \, d\mass ~.
\end{equation}
A useful consequence is
\eq{
\left.\frac{\partial \killing_c}{\partial T}\right|_{D_c}=-4\pi e^{Q_c}\frac{w_h'}{w_h''}\,.
}{eq:td143}
Holding the temperature $T_c$ fixed implies a relationship between $d \mass$ and $d D_c$, which can be obtained using the definition \eqref{Tc}, and the relations \eqref{dX_and_dT} and \eqref{dAc}:
\begin{equation}\label{dD_T}
  \left( \frac{1}{2\pi}\,\frac{\partial T}{\partial X_h} + e^{Q_c}\,T_c^{\,2} \right) d\mass \Big|_{T_c} = \frac{1}{2}\,T_c^{\,2} \,\left( U_c \, \killing_c + e^{Q_c}\,w'_c \right) d D_c \Big|_{T_c} ~.
\end{equation}
Using \eqref{FreeEnergy} and the formulas above we obtain
\eq{
\left.\frac{\partial \psi_c}{\partial D_c}\right|_{T_c} = -\left(\iscoa_c+\iscob_c\sqrt{\killing_c}+\iscoc_c\frac{1}{\sqrt{\killing_c}}\right)
}{eq:td111}
with
\eq{
\begin{aligned}
\iscoa &= \frac{e^{-Q/2}}{2\sqrt{w}}\left[w''-\frac{{w'}^2}{2w}-(wU)'+\frac12 wU^2\right]\\
\iscob &= \frac12 e^{-Q} \left(U'-U^2+U N\right)\\
\iscoc &= -\frac12 w''+\frac12 w' N
\end{aligned}
}{eq:lala}
where
\eq{
N = -w''_h\,\frac{(2M-w)U-w'}{{w'_h}^2-2w''_h(2M-w)}\,.
}{eq:td112}
The relations \eqref{eq:td143} and
\eq{
\left.\frac{\partial \killing_c}{\partial T_c}\right|_{D_c}=\left.\frac{\partial \killing_c}{\partial T}\right|_{D_c}\sqrt{\killing_c}\left(1-\frac{T}{2\killing_c}\left.\frac{\partial \killing_c}{\partial T}\right|_{D_c}\right)^{-1}
}{eq:td142}
allow to determine
\eq{
\left.\frac{\partial \psi_c}{\partial T_c}\right|_{D_c}=-\frac{1}{4}\,\killing_c^{-3/2}\left(w'_c+\killing_c U_c e^{-Q_c}\right)\left.\frac{\partial \killing_c}{\partial T_c}\right|_{D_c}\,.
}{eq:td141}
Another useful formula is
\eq{
\left.T_c \frac{\partial S}{\partial T_c}\right|_{\psi_c}=-T_c\frac{\left.\frac{\partial \psi_c}{\partial T_c}\right|_S}{\left.\frac{\partial \psi_c}{\partial S}\right|_{T_c}}\,.
}{eq:useful}

\providecommand{\href}[2]{#2}\begingroup\raggedright\endgroup

\end{document}